\documentclass[10pt,twocolumn,twoside]{IEEEtran}
\usepackage{cite}

\ifCLASSINFOpdf
  \usepackage[pdftex]{graphicx}
\else
\fi
\usepackage{amsmath}
\usepackage{algorithmic}
\ifCLASSOPTIONcompsoc
  \usepackage[caption=false,font=normalsize,labelfont=sf,textfont=sf]{subfig}
\else
  \usepackage[caption=false,font=footnotesize]{subfig}
\fi

\usepackage{tabu}
\usepackage{booktabs}
\usepackage{makecell}
\usepackage[pagebackref=true,breaklinks=true,letterpaper=true,colorlinks,bookmarks=false]{hyperref}
\usepackage{algorithm}
\usepackage[table,x11names]{xcolor}
\usepackage{amsfonts,amssymb,amsbsy}
\usepackage{multirow}
\usepackage{enumitem}
\usepackage{tabularx}

\newcolumntype{L}[1]{>{\raggedright\arraybackslash}p{#1}}
\newcolumntype{C}[1]{>{\centering\arraybackslash}p{#1}}
\newcolumntype{R}[1]{>{\raggedleft\arraybackslash}p{#1}}
\definecolor{mygreen}{HTML}{55d400}

\begin{document}
%
\title{Multi--Grid Back--Projection Networks}
%
%
%

\author{
Pablo Navarrete~Michelini,~\IEEEmembership{Member,~IEEE,}
Wenbin Chen,
Hanwen Liu,
Dan Zhu,
Xingqun Jiang
\thanks{Pablo Navarrete M., Wenbin Chen, Hanwen Liu, Dan Zhu, and Xingqun Jiang are with the Department
of Artificial Intelligence, BOE Technology Group Co., Ltd., Beijing,
China.}
\thanks{Manuscript received July $1^{st}$, 2020. First revision November $17^{th}$, 2020. Second revision December $23^{rd}$, 2020. Accepted December $28^{th}$, 2020.}}

%
%

\markboth{}%
{}
%



\maketitle

\begin{abstract}
Multi--Grid Back--Projection (MGBP) is a fully--convolutional network architecture that can learn to restore images and videos with upscaling artifacts. Using the same strategy of multi--grid partial differential equation (PDE) solvers this multiscale architecture scales computational complexity efficiently with increasing output resolutions. The basic processing block is inspired in the iterative back--projection (IBP) algorithm and constitutes a type of cross--scale residual block with feedback from low resolution references. The architecture performs in par with state--of--the-arts alternatives for regression targets that aim to recover an exact copy of a high resolution image or video from which only a downscale image is known. A perceptual quality target aims to create more realistic outputs by introducing artificial changes that can be different from a high resolution original content as long as they are consistent with the low resolution input. For this target we propose a strategy using noise inputs in different resolution scales to control the amount of artificial details generated in the output. The noise input controls the amount of innovation that the network uses to create artificial realistic details. The effectiveness of this strategy is shown in benchmarks and it is explained as a particular strategy to traverse the perception--distortion plane.
\end{abstract}

\begin{IEEEkeywords}
multigrid, iterative backprojection, super resolution, convolutional networks.
\end{IEEEkeywords}

%
\IEEEpeerreviewmaketitle

%
%
%
%

\section{Introduction}
\label{sec:introduction}
\begin{figure}[b]
  \centering
  \includegraphics[width=\linewidth]{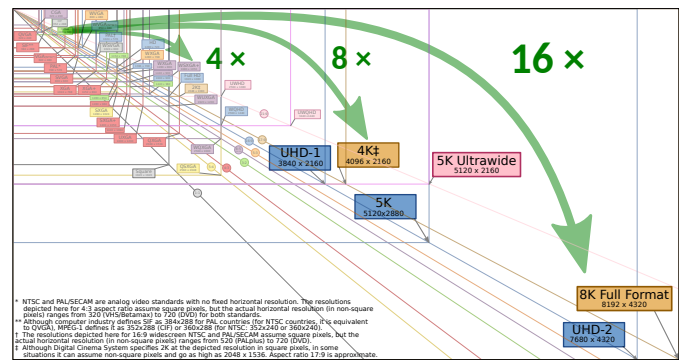}
  \caption{The dramatic growth of standard resolutions demands the development of Super--Resolution technologies to resize image with large upscaling factors. Source: \href{https://commons.wikimedia.org/wiki/File:Vector_Video_Standards.svg}{Wikimedia CC BY-SA}.\label{fig:standards}}
\end{figure}

\IEEEPARstart{I}{mage} and video upscaling has been studied for decades and remains an active topic of research because of constant technological advances in digital imaging.
One scenario where upscaling is now more demanding arises in digital display technologies, where new standards like BT.2020~\cite{MSugawara_2014a} are introduced. The resolution of digital displays has experienced a tremendous growth over the past few decades, as shown in Figure \ref{fig:standards}. The transition between different formats leads to a challenging problem. On one hand, large amount of digital content still exist in popular old standards such as standard--definition (SD). On the other hand, the latest display technologies (e.g. 4K, 8K and above) are expected to show this content with reasonable quality. Standard upscaling technologies are clearly insufficient for this purpose. While a $2\times$ upscaler maps $1$ input pixel into $4$ output pixels, an $8\times$ upscaler maps $1$ input pixel into $64$ output pixels, which already can contain a small image. The problem demands advanced solutions that are capable of understanding the content and filling in these large pieces of images with visually appealing and consistent information. In particular, large upscaling factors are needed to convert SD to ultra high--definition (UHD) resolutions. Thus, large upscaling represents a real problem in current market and it is expected to persist and become even more challenging with the rapid development of new technologies.

In classical interpolation theory, upscaling images by integer factors is explained as two sequential processes: upsample (insert zeros) and filter \cite{JGProakis_2007a,SMallat_1998a}. Standard upscaler algorithms, such as Bicubic or Lanczos, find high--resolution images with a narrow frequency content by using fixed low--pass filters. Modern tensor processing frameworks (e.g. Pytorch, Tensorflow, etc.) implement this process using a so--called \emph{strided transposed convolutional layer}. Similarly, the image acquisition process can be modeled as: filter and downsample (drop samples). Many times we know the explicit model, e.g. bicubic downscaler. Tensor processing frameworks implement this process using a \emph{strided convolutional layer}.

More advanced upscalers have followed geometric principles to improve image quality. For example, \emph{edge--directed interpolation} uses adaptive filters to improve edge smoothness \cite{VRAlgazi_1991a,XLi_2001a}, or \emph{bandlet} methods use both adaptive upsampling and filtering \cite{SMallat_2007a}. Later on, machine learning has been able to use examples of pristine high--resolution images to learn a mapping from low--resolution \cite{SCPark_2003a}. The rise of deep--learning and convolutional networks in image classification tasks \cite{YLeCun_2015a} quickly saw a series of important improvements in image resizing with large upscaling factors, which is the process widely know as image super--resolution (SR). Major progress in network architectures for image classification often succeeded in image SR, as seen for example with CNNs applied in SRCNN~\cite{Dong_2014a}, ResNets~\cite{he2016deep} applied in EDSR~\cite{Lim_2017_CVPR_Workshops}, DenseNets~\cite{huang2017densely} applied in RDN~\cite{zhang2018residual}, attention~\cite{hu2018squeeze-and-excitation} applied in RCAN~\cite{zhang2018rcan}, and non--local attention~\cite{wang2018non-local} applied in RNAN~\cite{zhang2019residual}. In all these examples, arguably the most influential practice is the use of residual networks (ResNets).

For video content, single frame upscaling would loose the chance of improving quality by learning temporal correlations and can result in visible flickering artifacts at large upscaling factors. Popular approaches to capture temporal information include: methods that align neighbor frames in addition to solve single image artifacts \cite{farsiu2004fast,liu2014on,ma2015handling}, network architectures that use motion compensation to align neighbor frames \cite{kappeler2016video,liao2015video,liu2017robust,tao2017detail}, use transformer blocks to learn the optical flow \cite{xue2019video}, learn dynamic upscaling filters \cite{jo2018deep}, and more recently use deformable convolutions for frame alignment \cite{wang2019edvr}. To better understand the mechanisms used to extend image to video SR we describe some of these solution in more detail:

\begin{itemize}[leftmargin=*]
\item TOFlow~\cite{xue2019video} is a network designed with motion estimation and video processing components. It combines three modules to: 1) estimate the motion fields between input frames; 2) register all input frames based on estimated motion fields; and 3) generate target output from registered frames. These three modules are jointly trained to minimize the loss between output frames and ground truth.

\item Dynamic Upsampling Filters (DUF)~\cite{jo2018deep} is a network that generates dynamic upsampling filters and a residual image, which are computed depending on the local spatio--temporal neighborhood of each pixel to avoid explicit motion compensation. A HR image is reconstructed directly from the input image using the dynamic upsampling filters, and the fine details are added through the computed residual.

\item Enhanced Deformable convolutions Video Restoration (EDVR)~\cite{wang2019edvr} is a network that combines frame alignment and spatio--temporal fusion. To handle large motions it uses a Pyramid, Cascading and Deformable (PCD) alignment module, in which frame alignment is done at the feature level using deformable convolutions in a coarse--to--fine manner. Then they use a Temporal and Spatial Attention (TSA) fusion module, in which attention is applied both temporally and spatially to emphasize important features for subsequent restoration.

\item The Video Enhancement and Super--Resolution network (VESR--Net)~\cite{chen2020vesr} uses both PCD alignment and a Separate Non--Local attention (SNL) module to aggregate the information among different frames. For reconstruction, they utilize stacked channel--attention residual block (CARB)~\cite{zhang2018image} followed by a feature decoder. Finally, Efficient Video Enhancement and Super--Resolution Net (EVESR--Net)~\cite{fuoli2020aim} improved VESR--Net replacing the SNL module with an Efficient Point-Wise Temporal Attention Block (EPAB) that aggregates the spatio--temporal information with less operations and memory consumption.
\end{itemize}

The optimization problem of learning to map images/videos from low to high resolution can have two distinctive targets:
\begin{itemize}[leftmargin=*]
    \item \textbf{High Fidelity}, the aim is to obtain a solution capable to produce high resolution results with the best fidelity (PSNR) to the ground truth. This is also known as a \emph{low distortion} target where \emph{distortion} refers to a measure of the difference between output content and ground truth.
    \item \textbf{Perceptual Quality}, the aim is to obtain a solution capable to produce high resolution results with the best perceptual quality similar to the ground truth.
\end{itemize}
Historically, high fidelity was the major focus of research in image SR until systems became more efficient and large upscaling factors where considered. The empirical evidence then showed that high fidelity often leads to cartoonish effects that are easily recognizable and looked far from real photographs. Soon this reveal a fundamental phenomena that is now known as the \emph{Perception--Distortion} trade--off established by Blau and Michaeli in \cite{Blau_2018_CVPR}. That is, both perception and distortion targets cannot be achieved at the same time, one must compromise perceptual quality to improve fidelity and vice versa. The trade--off also motivated the first workshop and challenge focused on perceptual super--resolution (PIRM 2018~\cite{blau20182018}) that strongly confirmed and gave further insight into this principle.

Here, we present our contributions to the development of new image and video super--resolution technologies with major focus on: deep--learning architectures, and training strategies for perceptual super--resolution. First, we propose a deep learning architecture called Multi--Grid Back--Projection (MGBP)\cite{PNavarrete_2019a, G-MGBP, MGBPv2} that has the following important characteristics:
\begin{itemize}[leftmargin=*]
    \item It uses a \emph{back--projection} residual strategy that moves features from high to low resolution to compare them with a low resolution reference and then comes back to correct the high resolution feature. This is analogous to the iterative back--projection (IBP) algorithm but its target is more far reaching as a general learning module that aims to improve the high resolution output in successive iterations.
    \item It uses a \emph{multi--grid} recursive strategy using back--projection residual blocks within back--projection residual blocks. Thus, it quickly moves features to the lowest resolution stages where it concentrates most of the heavy processing and leaves easier tasks for higher resolutions, efficiently balancing the computational complexity across scales. This is analogous to multigrid PDE solvers that are known to be optimal in the solution of large systems of equations \cite{UTrottenberg_2000a}.
    \item It allows a recursive configuration sharing parameters in all back--projection blocks leading to good results and very small number of parameters, or alternatively, a non--recursive configuration that leads to better performance and an increased number of parameters.
    \item It can be extended to use 3D--convolutions for video. This is due to the property that back--projection blocks do not change the size of the input and allows to learn a mapping from many input to many output frames. This is a convenient property that we call \emph{Cube--to--Cube} and it leads to a fully 3D--convolutional network that we call MGBP--3D. It departs significantly from other video solutions mentioned above as it relies completely on 3D--convolutions and do not include other non--linear blocks such as attention, warping, dynamic filters or deformable convolutions.
\end{itemize}
Second, we designed a particular training strategy for perceptual super--resolution with the following contributions:
\begin{itemize}[leftmargin=*]
    \item We propose a strategy to control the perception--distortion trade-off as a transition from a high--fidelity target to perceptual target through a so called \emph{innovation jump}.
    \item We drive the innovation process of our MGBP architecture (generator) by introducing random inputs at each resolution level. These inputs are manipulated by the network to generate artificial details at different scales.
    \item We propose a so--callled \emph{variance--normalization and shift--correlator} (VN+SC) layer that provides meaningful features to a discriminator system based upon previous research on the statistics of natural images.
    \item We propose a \emph{multiscale discriminator} for adversarial training. It is a configuration symmetric to the multi--scale upscaler, therefore it is more effective for adversarial training. It can simultaneously evaluate several upscaling factors, resembling a Progressive GAN\cite{karras2018progressive} in the sense that the optimizer can focus on smaller factors first and then work on larger factors.
    \item We propose a novel \emph{noise--adaptive training strategy} that can avoid conflicts between reconstruction and perceptual losses, combining loss functions with different random inputs.
\end{itemize}

An early version of the MGBP architecture first appeared in the NTIRE 2018 challenge on image SR\cite{NTIRE2018_SR_Report}, together with Deep Back--Projection Networks (DBPN)\cite{DBPN2018}. Both DBPN and MGBP used similar back--projection residual blocks with DBPN achieving better results and winning this challenge with a two--level configuration mixing up/down--backprojections. The architecture of MGBP was later improved in \cite{PNavarrete_2019a} achieving state--of--the--arts results in high--fidelity SR for lightweight networks with small number of parameters. The training strategy for perceptual quality first appeared in PIRM 2018 achieving the $\boldsymbol{2^{nd}}$ \textbf{best perceptual quality} in the Perceptual SR challenge \cite{blau20182018}. We refer to these early configurations using small number of parameters as MGBP version 1 (MGBPv1). The architecture was later improved in \cite{lugmayr2019aim} to allow a non recursive structure that is more flexible and simple to configure, allowing a larger amount of parameters while at the same time being able to process very large images (e.g. 8K). This configuration was named MGBP version 2 (MGBPv2) and has won the $\boldsymbol{1^{st}}$ \textbf{place for best perceptual quality} in the AIM 2019 challenge for Extreme Super--Resolution ($16\times$ factor) as well as $\boldsymbol{3^{rd}}$ \textbf{best perceptual quality} in NTIRE 2020 challenge on Real--World Image SR. Here, we include new experiments to show that MGBPv2 can also achieve state--of--the--arts results in high--fidelity targets for image SR. Finally, we show how to extend MGBPv2 to the fully 3D--convolutional network MGBP--3D. This new architecture was recently used in the AIM 2020 Video Extreme Super--Resolution Challenge achieving the $\boldsymbol{2^{nd}}$ \textbf{best fidelity} \cite{fuoli2020aim}.
Overall, the MGBP architecture has been particularly successful as a convenient configuration to balance multi--scale computational complexity for large upscaling factors in both image and video SR as well as achieving state--of--the--arts results in perceptual super--resolution.

\begin{figure}
    \centering
    \includegraphics[width=0.9\linewidth]{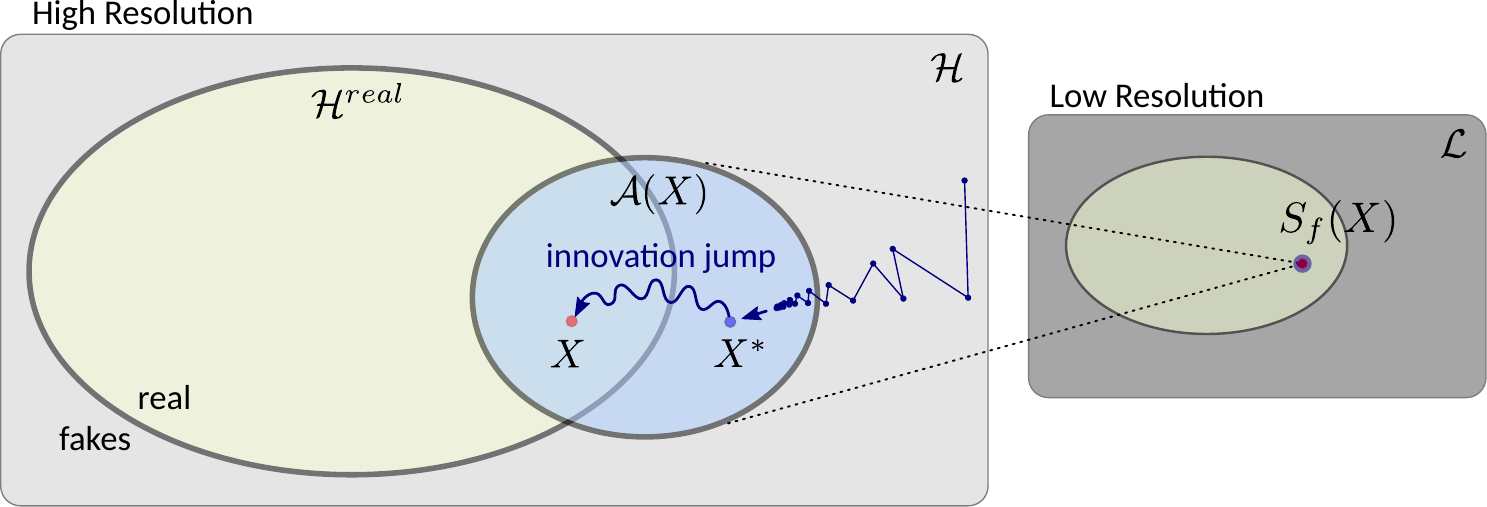}
    \caption{For a high--resolution image $X$ that looks real, distortion optimization approaches an optimal solution $X^*$ that does not look real because it lacks the unpredictable nature of natural images. We can still use $X^*$ as a reference point to move through an \emph{innovation jump} into the set of realistic images. \label{fig:strategy_sets}}
\end{figure}

\begin{figure*}
  \centering
  \includegraphics[width=\linewidth]{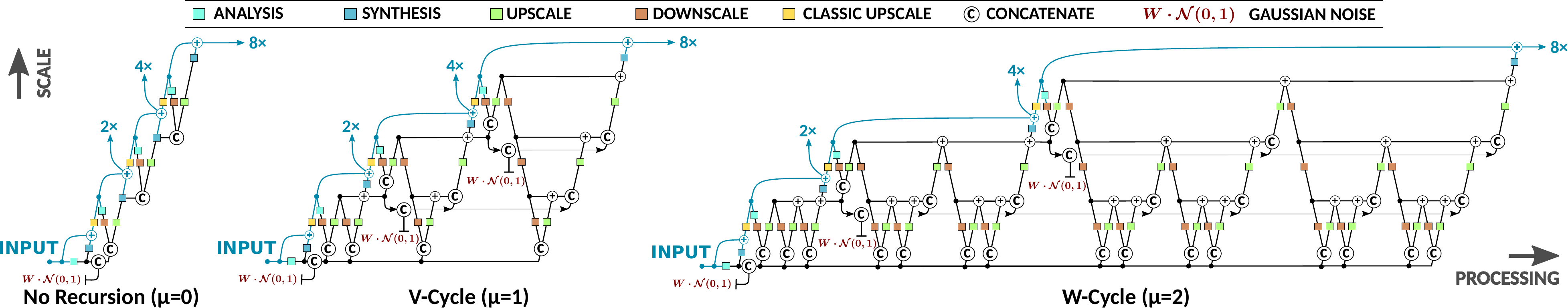}
  \caption{Diagram of the MultiGrid BackProjection version 1 (MGBPv1) network \cite{PNavarrete_2019a} with $\mu=0$, $1$, and $2$. \label{fig:mgbpv1}}
\end{figure*}
\begin{figure*}
  \centering
  \includegraphics[width=\linewidth]{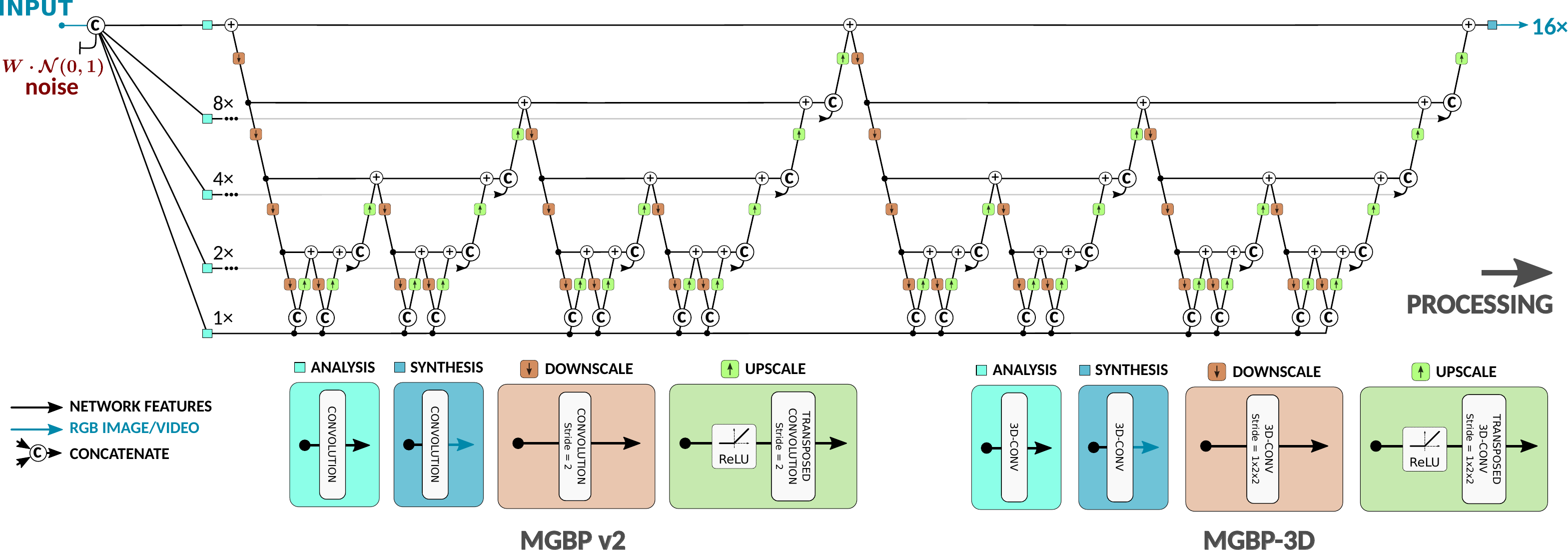}
  \caption{Diagram of the MultiGrid BackProjection network version 2 (MGBPv2) for images and MGBP--3D for videos unfolded from Algorithm \ref{alg:mgbp2} with $\mu=2$. \label{fig:mgbpv2}}
\end{figure*}

\section{Perceptual and Distortion Targets}
\label{sec:strategy}

To better illustrate our target, we present a diagram of image sets in Figure \ref{fig:strategy_sets}. Here, $\mathcal{H}$ is the set of all high--resolution images, $\mathcal{H}^{real}\subset\mathcal{H}$ is the subset of high--resolution images that correspond to natural images, and $\mathcal{L}$ is the set of all low--resolution images. Given an image $X\in\mathcal{H}^{real}$, we are interested in the set of \emph{aliased} images:
\begin{equation}
    \mathcal{A}(X) = \left\{ Y \in \mathcal{H} \quad s.t. \quad S_f(Y)=S_f(X) \right\} \;,
\end{equation}
where $S_f:\mathcal{H}\rightarrow\mathcal{L}$ is a \emph{downscale} operator of factor $f$. We are particularly interested in the set $\mathcal{A}(X)\cap\mathcal{H}^{real}$ of alias images that correspond to real content.

A \emph{distortion} function $\Delta(X,y)$ measures the dissimilarity between a reconstructed image $y$ and the original image $X$. Popular and basic distortion metrics such as L1, L2, PSNR, etc., are sensitive to changes (any minor difference in pixel values would increase the amount of distortion) and are known to have low correlation with human perception\cite{KSeshadrinathan_2010a}. Several distortion metrics have been proposed to approach perceptual quality by emphasizing some differences more than others, either through normalization, feature extraction or other approaches. These include metrics like SSIM\cite{Wang04imagequality}, VIF\cite{VIF} and the VGG content loss\cite{DBLP:journals/corr/JohnsonAL16}. By doing so, correlation with human perception improves according to \cite{KSeshadrinathan_2010a}, but experiments in \cite{Blau_2018_CVPR} show that these metrics still focus more on distortion. More recently, the contextual loss has been proposed to focus more on perceptual quality while maintaining a reasonable level of distortion\cite{mechrez2018contextual}.

The solution of distortion optimization is obtained by:
\begin{equation}
    X^*=\text{argmin}_y \mathbb{E}\left[\Delta(X,y)\right] \;. \label{eq:optimal_distortion}
\end{equation}
The original image $X$ is fixed, and the expected value in \eqref{eq:optimal_distortion} removes any visible randomness in the search variable $y$. But, according to research on the statistics of natural images, randomness plays an essential role in what makes images look real\cite{Ruderman1994}. This is well known for non--reference image quality metrics such as NIQE\cite{NIQE_2013} or BRISQUE\cite{BRISQUE_2012}, and led to a definition of perceptual quality as a distance between probability distributions in \cite{Blau_2018_CVPR}. It is also known that distortion optimization solutions tend to look unreal, as seen in state--of--the--art results from NTIRE--SR Challenges\cite{Timofte_2017_CVPR_Workshops,NTIRE2018_SR_Report}. Common distortion metrics in these challenges (L1 and L2) make the image $X^*$ lose all randomness. We argue that this removal of randomness in $X^*$ is what moves it out of set $\mathcal{H}^{real}$, as we show in Figure \ref{fig:strategy_sets}.

We know that $X\neq X^*$ because $X\in\mathcal{H}^{real}$ and $X^*\notin\mathcal{H}^{real}$ according to our previous discussion. However, distortion optimization can still be useful to generate realistic images. By approaching $X^*$ we are getting closer to $X$. As shown in Figure \ref{fig:strategy_sets}, both $X$ and $X^*$ can be in $\mathcal{A}(X)$. Using a signal processing terminology, the \emph{innovation}\cite{Mitter_1982} is the difference between $X$ and the optimal forecast of that image based on prior information, $X^*$. Most SR architectures take the randomness for the innovation process from the low--resolution input image, which is a valid approach but loses the ability to expose and control it.

In our proposed architecture we add randomness explicitly as noise inputs, so that we can control the amount of innovation in the output. Independent and identically distributed noise will enter the network architecture at different scales, so that each of them can target artificial details of different sizes. Generally speaking, our training strategy will be to approach $X^*$ with zero input noise and any image in $\mathcal{A}(X)\cap\mathcal{H}^{real}$ with unit input noise. By using noise to target perceptual quality, and remove it for the distortion target, we teach the network to \emph{jump} from $X^*$ into $\mathcal{H}^{real}$. With probability one the network cannot hit $X$, but the perceptual target is any image in $\mathcal{A}(X)\cap\mathcal{H}^{real}$.

\begin{algorithm*}
    \caption{Multi--Grid Back--Projection version 2 (MGBPv2)} \label{alg:mgbp2}
    \begin{tabular}{ll}
        \textbf{MGBPv2}$(X,W,\mu,L)$: & $\boldsymbol{BP^{\mu}_{k}}(u\;|\;y_1,\ldots,y_{k-1}; \text{tag}_1,\ldots,\text{tag}_{k-1})$: \\

        \resizebox{.45\textwidth}{!}{
        \begin{minipage}{.5\textwidth}
            \begin{algorithmic}[1]
                \REQUIRE Input image $X$.
                \REQUIRE Steps $\mu$, levels $L$ and noise amplitude $W$.
                \ENSURE Output image $Y$.

                \STATE $x = [X, \; W\cdot\mathcal{N}(0,1)]$
                \FOR{$k = 1,\ldots,L$}
                    \STATE $y_k = \text{Analysis}_{k}(x)$
                \ENDFOR
                \FOR{$k = 1,\ldots,L$}
                    \STATE $\text{tag}_k = 0$
                \ENDFOR
                \STATE $y = BP^{\mu}_L(y_L\;|\;y_1,\ldots,y_{L-1}; \text{tag}_1,\ldots,\text{tag}_{L-1})$
                \STATE $Y = \text{Synthesis}(y)$
            \end{algorithmic}
        \end{minipage}
        }

        &

        \resizebox{.45\textwidth}{!}{
        \begin{minipage}{0.53\textwidth}
            \begin{algorithmic}[1]
                \REQUIRE Input image $u$, level index $k$, steps $\mu$.
                \REQUIRE Images $y_1,\ldots,y_{k-1}$ and $\text{tag}_1,\ldots,\text{tag}_{k-1}$ (for $k>1$).
                \ENSURE Image $out$

                \STATE $out = u$
                \IF{$k > 1$}
                    \FOR{$s = 1,\ldots,\mu$}
                        \STATE $\text{tag}_{k-1} = s$
                        \STATE $LR = \text{Downscale}_{tag}(out)$
                        \STATE $c = BP^{\mu}_{k-1}(LR\;|\;y_1,\ldots,y_{k-2}; \text{tag}_1,\ldots,\text{tag}_{k-2})$
                        \STATE $out = out + \text{Upscale}_{tag}([\;y_{k-1}, c\;])$
                    \ENDFOR
                \ENDIF
            \end{algorithmic}
            \end{minipage}
        }
    \end{tabular}
\end{algorithm*}

\section{Architecture}
\subsection{Generator}
\textbf{Motivation}. The general problem of image restoration is to recover an image $X$ that has been degraded. This could mean that either some details have been removed from the image (e.g. blurring) or some new information is interfering and corrupting the image (e.g. noisy measurements). Here, we will focus on the problem of image super--resolution, assuming a degradation model of the form:
\begin{equation}
    x = R(X) \;, \label{eq:model_down}
\end{equation}
where $X\in\mathbb{R}^N$ is the high--resolution source, $x\in\mathbb{R}^n$ is the low--resolution result, and $R:\mathbb{R}^N\rightarrow\mathbb{R}^n$ is a \emph{restriction} operator (downscaler). A classic linear model is $R(X) = (X*g)\downarrow s$ where $g$ is a blurring kernel (e.g. bicubic) and $\downarrow s$ is a downsampling by factor $s$ dropping pixels to reduce the resolution.

The degradation model \eqref{eq:model_down} represents a prior knowledge that we would like to enforce on our system, aiming to recover the original image. This is the motivation behind the classic Iterative Back--Projection (IBP) algorithm \cite{Irani_1991a}. Given model \eqref{eq:model_down} and an upscaled image $y_0$, the IBP algorithm iterates:
\begin{align}
    e_k     & = x - R(y_k) \;, \label{eq:error}\\
    y_{k+1} & = y_k + P(e_k) \;. \label{eq:correction}
\end{align}
Here, $e(y_k)$ is the mismatch error at low--resolution and $P:\mathbb{R}^n\rightarrow\mathbb{R}^N$ is a projection operator (upscaler). A classic linear projection is $P(x)= (x\uparrow s) * p$, where $p$ is an upscaling filter (a convolution) and $\uparrow s$ is increasing the resolution by inserting zeros. For linear operators $P$ and $R$, it is simple to see that $y_k$ will follow the degradation model \eqref{eq:model_down} as $k\rightarrow\infty$ when $||I-RP||<1$ \cite{Irani_1991a}.

IBP computes a residual $P(e_k)$ to update the output in successive iterations. It is thus natural to think of this process as a residual block. The restriction operator $R$ can be implemented with a strided convolutional layer and the projection operator $P$ with a strided transposed convolution. In MGBP we use these residual steps and follow the design of residual blocks in EDSR~\cite{Lim_2017_CVPR_Workshops} by using a single rectified linear unit just before the transposed convolution in the $P$ operator. We also make a small generalization by changing the error equation \eqref{eq:error} to $e_k = [x, R(y_k)]$ where $[\cdot, \cdot]$ replaces the difference to concatenation of features (in the channel dimension). The projector operator $P$ will then decide how to compare $x$ and $R(y_k)$ in the update equation \eqref{eq:correction}.

The IBP iteration does work as a residual block but it is different to the conventional blocks used in ResNets~\cite{he2016deep} in that IBP performs the correction at a lower resolution. This is similar to the way a Full--Multigrid algorithm solves linear equations \cite{UTrottenberg_2000a}. That is: first, find an approximate solution in a small grid; and second, interpolate the approximate solution to update the solution in a large grid. Here, we borrow an essential idea of multigrid methods, that is to repeat the process recursively with residual backprojection blocks within residual backprojection blocks that quickly move features the lowest resolutions. Hence, two step back--projection residuals in high resolution ($\mu=2$) leads to $4$ residual steps in the next lower resolution, $8$ in the next level, and so forth. The pattern unfolded is known in multigrid literature as a W--cycle\cite{UTrottenberg_2000a} as seen in Figure \ref{fig:mgbpv1}.

The MGBP algorithm has been updated from version 1 (MGPBv1) in \cite{PNavarrete_2019a} to version 2 (MGPBv2) in \cite{MGBPv2} but the core idea explained above has remained unchanged. The latest Multi--Grid Back--Projection algorithm (MGBPv2) is shown in Figure \ref{fig:mgbpv2} that follows from Algorithm \ref{alg:mgbp2}.

\begin{figure}
    \centering
    \includegraphics[width=\linewidth]{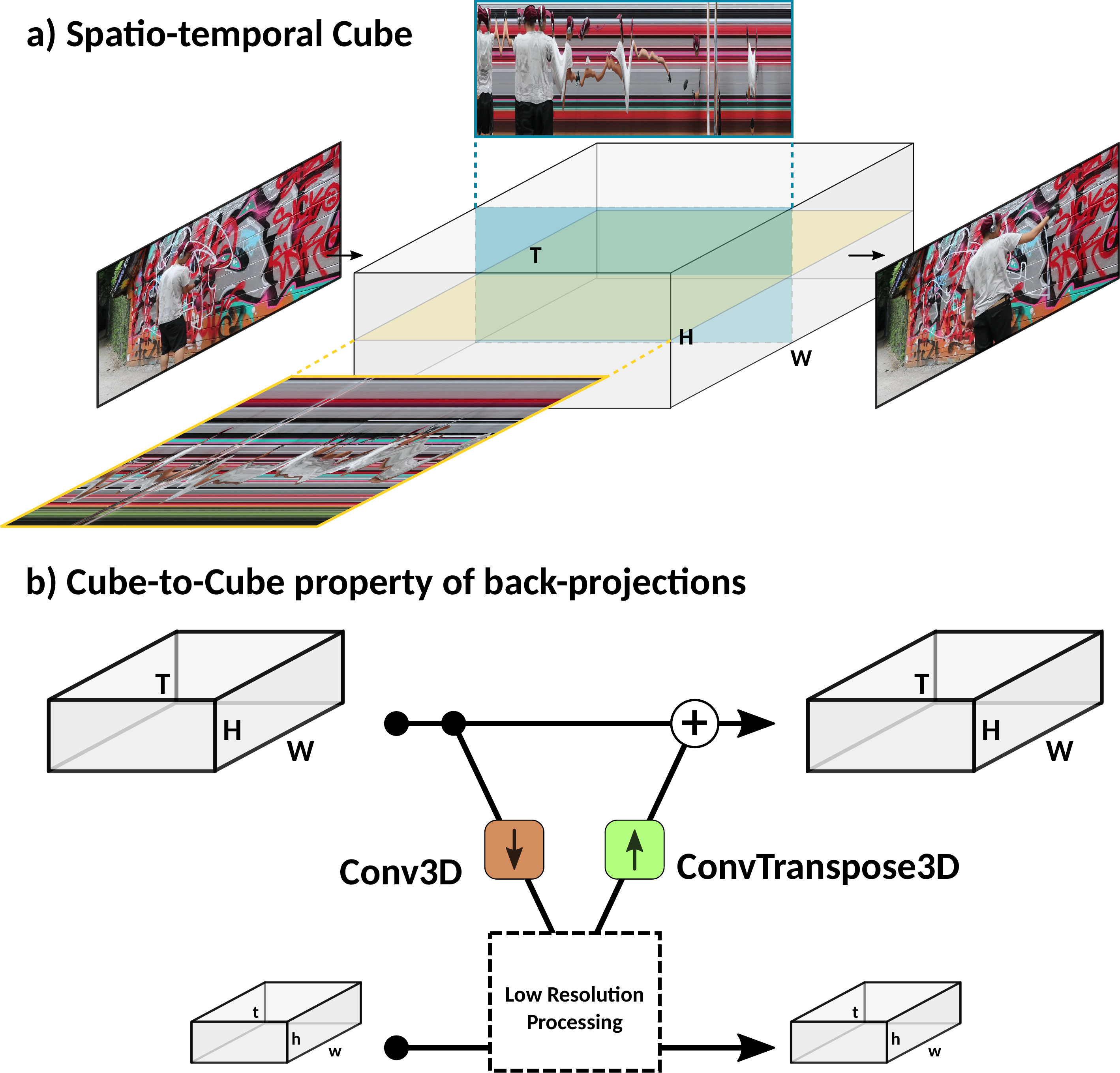}
    \caption{a) A video clip is represented as a spatio-temporal cube. b) A back--projection module using a convolution and transposed convolution with same kernel size preserve the dimensions of the cube. Padding is no needed and can reduce the memory footprint at low resolution levels. \label{fig:cube}}
\end{figure}

\textbf{Cube--to--Cube property}. When we upscale sequences of images that come from video streams it is well known that temporal information plays a crucial role. Independently super--resolving frames looses important temporal correlations and introduces temporal artifacts such as flickering. A simple model to add this information is to interpret time as an extra dimension. In this interpretation flickering artifacts are the temporal version of jaggies in space. We can visualize the spatio--temporal information of a video clip in a cubical arrangement of pixels as seen in Figure \ref{fig:cube}--a. The cross sections of the cube along temporal dimension reveal a smooth continuity of pixel value changes across time. A common solution for video super--resolution is to use 3D convolutions to extract this information based on the fact that these cross sections share strong similarities with 2D images (eg. sharp and smooth edges).

Extending a network model from 2D to 3D convolutions is particularly simple when using back--projections because of the property shown in Figure \ref{fig:cube}--b. As we already know, a back--projection module uses a strided convolution to move down in scale and strided transposed convolution to move back. Here, we assume that both convolution and transposed convolutions use the same kernel sizes and padding settings. Because the matrix representation of this transposed convolution has the same dimensions of the transposed matrix for its convolution counterpart, the residual update in the back--projection module has the exact same resolution as the initial state. This remains true with or without padding in the convolutional modules. We call this the \emph{Cube--to--Cube} property of back--projection modules. The property holds for any number of dimension (e.g. using 1D, 2D or 3D convolutions) but it is particularly useful for higher dimensions. This is because by removing padding we save a fix amount of pixels in 1D, a fix amount of lines (rows or columns) in 2D, and a fix amount of images in 3D, without reducing the receptive field of convolutional layers. The memory saving becomes significant when processing video as we move down in scale using back--projection blocks. For example, we used $37$ frames to train a $1\times 16\times 16\times$ MGBP--3D (scaling in space and not in time) that became $29$ frames at the lowest resolution of the network when using kernel size $3$ in time and no padding, compared to $37$ frames with padding. The volume saving at the lowest resolution is $\sim 22\%$ which allows us to use a large number of features ($256$). This is a clear advantage of using a fully 3D--convolutional network structure.

\begin{figure}
    \centering
    \includegraphics[width=0.9\linewidth]{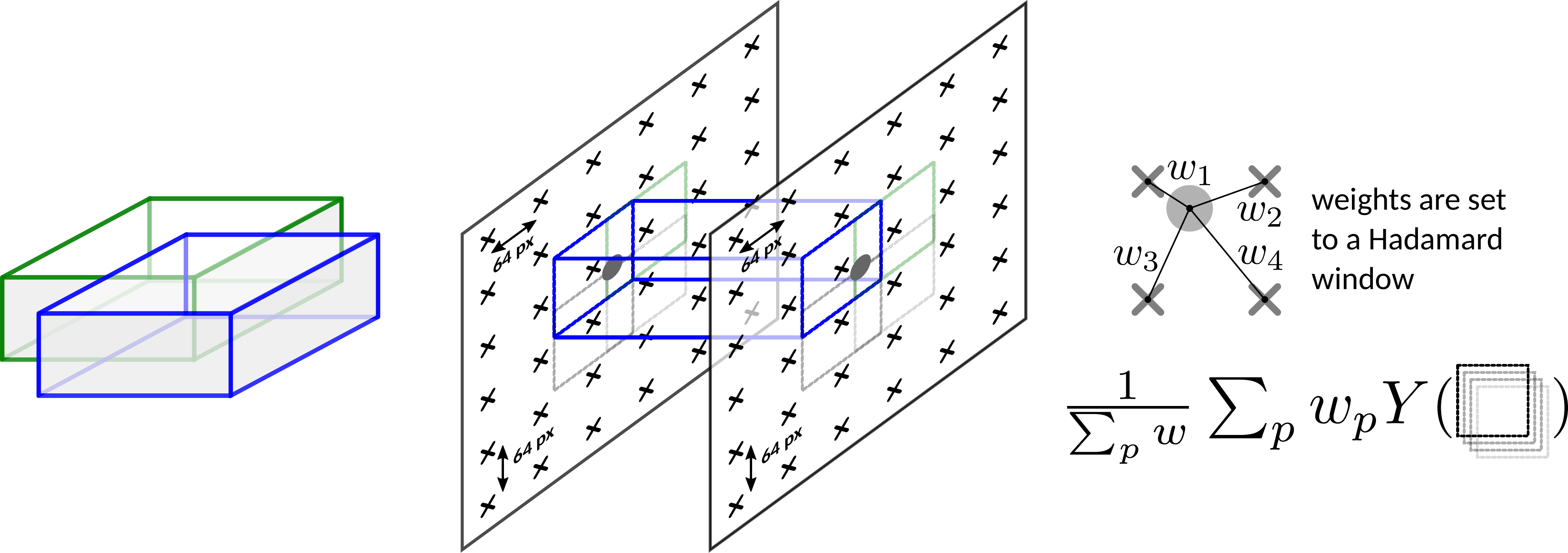}
    \caption{For video inference, using a fully 3D--convolutional architecture such as MGBP--3D is more difficult than image since in practice we are unable to input a 3D--cube with the entire video sequence. We solve this issue by using overlapping 3D--patches that are averaged using a Hadamard window. \label{fig:3d-overlap}}
\end{figure}

\textbf{Video Inference}. To upscale large video sequences we propose a patch based approach in which we average the output of overlapping video patches taken from the bicubic upscaled input. First, we divide input streams into overlapping patches (of same size as training patches) as shown in Figure \ref{fig:3d-overlap}; second, we multiply each output by weights set to a Hadamard window; and third, we average the results. In our experiments we use overlapping patches separated by $5$ video frames.

This strategy is necessary for video as we cannot input a spatio--temporal cube with the complete input sequence. This limitation of fully 3D--convolutional networks could be avoided if tensor processing frameworks implement \emph{streaming 3D--convolutions} where output frames come out as input frames arrive, using a FIFO structure equivalent to the implementation of audio filters.

\textbf{Generator Architecture}. Algorithm \ref{alg:mgbp2} refers to the second version of MGBP~\cite{MGBPv2}. One of the differences between the first and second versions is the initialization step. Without loss of generality, in the second version we tackled the image enhancement problem with an input resolution equal to the output resolution. For the super--resolution task we input the low resolution image upscaled using a bicubic method. This helps to make the system become more general for applications and simplifies the process of generating pairs of input/output patches during training. In contrast, the first version MGBPv1 shown in Figure \ref{fig:mgbpv1} started at the lowest resolution and progressively upscaled the input with a bicubic upscaler, entering the multigrid cycle at different resolutions.

Figure \ref{fig:mgbpv2} displays the diagram of the MGBPv2 algorithm unfolded for $\mu=2$ and $L=5$. The \emph{Analysis} and \emph{Synthesis} modules convert an image into feature space and vice--versa using single convolutional layers. The \emph{Upscaler} and \emph{Downscaler} modules are composed of single strided (transposed and conventional) convolutional layers. An important observation in Algorithm \ref{alg:mgbp2} is the use of a \texttt{tag} label to differentiate each \emph{Upscaler} and \emph{Downscaler} module. This simple trick was introduced in MGBPv2 and makes every \emph{Upscaler} and \emph{Downscaler} module different in terms of parameters and hyper--parameters. In particular, we can now set a different number of features in convolutional layers from low to high resolution levels, allowing large images to be processed at high resolutions by using small number of features and increase it at lower resolutions. The system can be initialized by a \emph{dry run} of Algorithm \ref{alg:mgbp2} where no computation is performed and modules are defined with their correspondent \texttt{tag} labels. The MGBPv1 system did not include this feature and it was force to use identical \emph{Upscaler} and \emph{Downscaler} in all instances of these modules. In this case a single convolutional layer would not give good results and $4$--layer DenseNets were used inside these modules \cite{PNavarrete_2019a}.

The same diagram for MGBPv2 in Figure \ref{fig:mgbpv2} represents the architecture of MGBP--3D and the same Algorithm \ref{alg:mgbp2} is used for video SR. The difference between MGBPv2 and MGBP--3D is in the configuration of the \emph{Analysis}, \emph{Synthesis}, \emph{Upscaler} and \emph{Downscaler} modules that now used 3D--convolutions as shown in Figure \ref{fig:mgbpv2}. This simple extension is due to the \emph{Cube--to--Cube} property explained above.

\textbf{Complexity}. The W--cycle pattern shown in Figure \ref{fig:mgbpv2} is unfolded when using $\mu=2$, and it is also a popular configuration in multigrid methods \cite{UTrottenberg_2000a}. Let $f(n)$ be the computational complexity of the residual backprojection module in a resolution of $n = H\times W$ pixels. For example, $f(n)=\mathcal{O}(n^{1/2})$ if we use the same \emph{Upscaler} and \emph{Downscaler} modules all over the network. Then, the number of products $p(n)$ at resolution $n$ using MGBP obeys the recurrence relation:
\begin{equation}
    p(n) = f(n) + \mu \cdot p(n/4) \;,
\end{equation}
assuming a downscaling by factor $2\times2$ (horizontal and vertical). The solutions of this recurrence are known as the master theorem for divide--and--conquer~\cite{cormen2009introduction}. In MGBPv1 we have $f(n)=\mathcal{O}(n^{1/2})$ and $\mu=2$, that leads to a computational complexity $p(n)=\mathcal{O}(n\log n)$. This is inconvenient in practice and it made it impossible, for example, to apply MGBPv1 in the Extreme--SR AIM 2019 challenge with upscaling factor $16\times$ and 8K output resolution. The complexity can be reduced to $\mathcal{O}(\log n)$ if we manipulate the \emph{Upscaler} and \emph{Downscaler} modules such that $f(n)=\mathcal{O}(n^{1/2-\epsilon})$ for some $\epsilon > 0$. This is easily achieved by both MGBPv2 and MGBP--3D by reducing the number of features per level in convolutional layers as a function of $n$.

In terms of memory requirements the resolution of feature maps decreases by factor $2\times 2=4$ at each scale (using stride $2$ in \emph{Upscaler} and \emph{Downscaler} modules). All versions of MGBP need to keep one copy of the feature map at each scale (similar to UNets) and the geometric sum of pixel counts from different scales leads to $\mathcal{O}(n)$. Therefore, MGBP memory footprint scales correctly by linearly increasing the memory requirements as the number of pixels increases.

\begin{figure*}
  \centering
  \includegraphics[width=.8\linewidth]{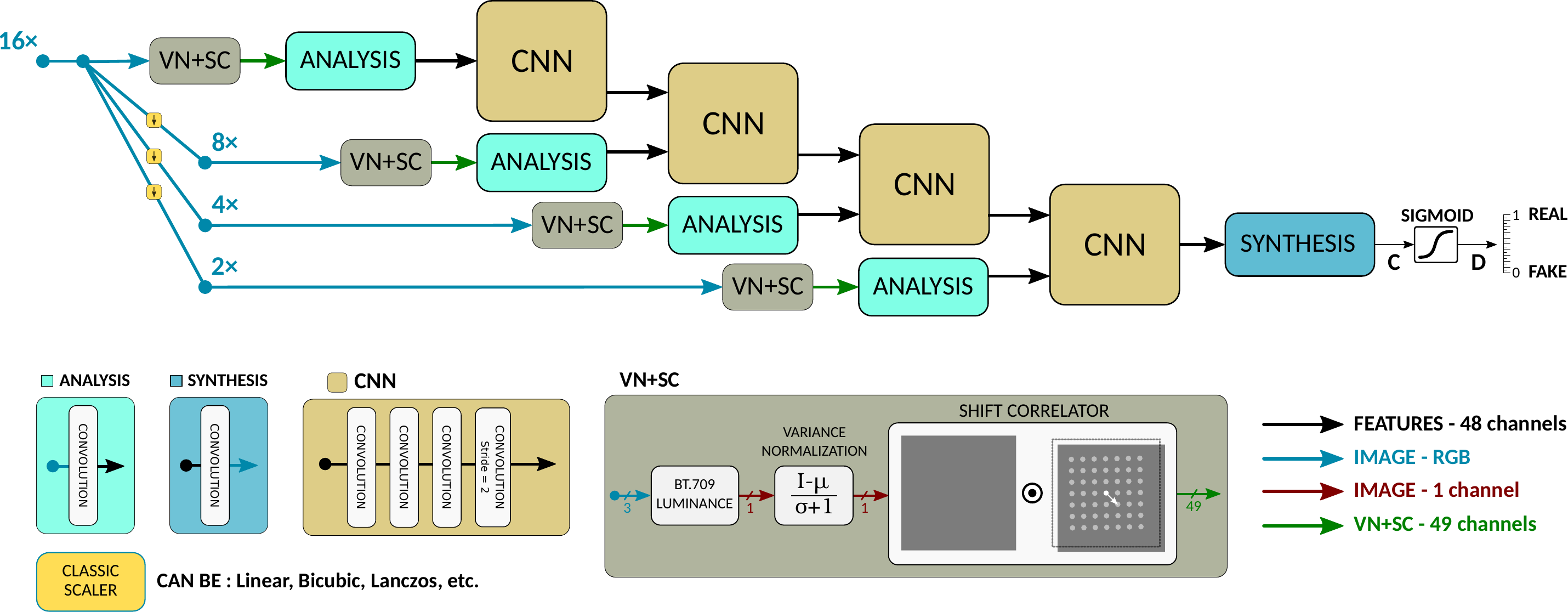}
  \caption{Discriminator system used for adversarial training. The high resolution input image is downsized with standard bicubic downscalers to enter the system at different scales. CNN modules do not share parameters. Only image content is considered for adversarial training as the temporal evolution of artificial details required for videos makes the problem significantly more difficult and currently unresolved. \label{fig:discr}}
\end{figure*}

\begin{figure}
    \centering
    \includegraphics[width=\linewidth]{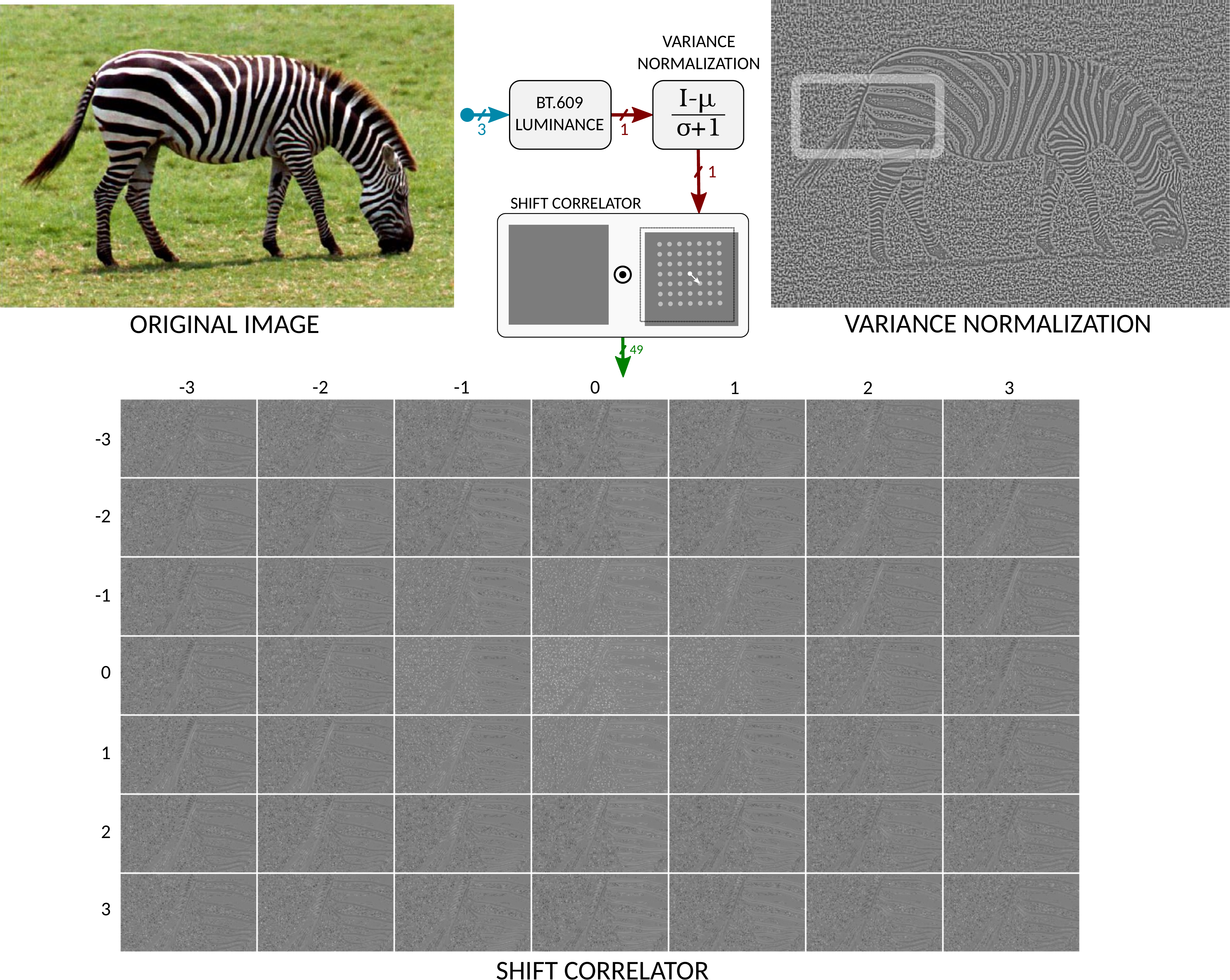}
    \caption{Our VN+SC layer considers the luminance channel of a color image (using the BT.609 standard for conversion from RGB), applies variance normalization (upper right image), and computes the Hadamard product with its shift versions (bottom image showing only outputs for the white box in variance normalization image). We use these $49$ channels as inputs of the Discriminator. \label{fig:vnsc}}
\end{figure}

\subsection{Discriminator}
The task of the discriminator is to measure how realistic is an image or video frame. A straightforward approach is to input the color image to a sequential convolutional network architecture. Then, we hope that the discriminator learns from adversarial training using real and fake image examples. In practice, we find that this approach works well to identify which areas of upscale images need more textures but the artificial details look noisy and have limited structure.

So what makes an image look natural? Extensive research has been carried to address this question. Here, we follow the seminal work of Ruderman\cite{Ruderman1994} who found regular statistical properties in natural images that are modified by distortions. In particular, Ruderman observed that applying the so--called \emph{variance normalization} operation:
\begin{equation}
    \hat{I}_{i,j} = \frac{I_{i,j}-\mu_{i,j}(I)}{\sigma_{i,j}(I)+1} \;,
\end{equation}
has a decorrelating effect on natural images. Here, $I_{i,j}$ is the luminance channel of an image with values in $[0, 255]$ at pixel $(i,j)$, $\mu(I)$ is the local mean of $I$ (e.g. output of a Gaussian filter), and $\sigma(I)^2=\mu(I^2)-\mu^2(I)$ is the local variance of $I$. Ruderman also observed that these normalized values strongly tend towards a Gaussian characteristic for natural images. These findings are used in the NIQE perceptual quality metric considered for the PIRM--SR Challenge 2018\cite{NIQE_2013}. NIQE also models the statistical relationships between neighboring pixels by considering horizontal and vertical neighbor products: $\hat{I}_{i,j}\hat{I}_{i,j+1}$, $\hat{I}_{i,j}\hat{I}_{i+1,j}$, $\hat{I}_{i,j}\hat{I}_{i,j-1}$ and $\hat{I}_{i,j}\hat{I}_{i-1,j}$.\smallskip

\textbf{Variance Normalization and Shift Correlator (VN+SC)}. Inspired by previous research we define the Variance Normalization and Shift Correlator (VN+SC) layer as follows:
\begin{equation}
    V^{7(p+3)+q+3}_{i,j}(I) = \hat{I}_{i,j}\cdot\hat{I}_{i+p,j+q} \;, \quad
    \begin{array}{ll}
        p=-3,\ldots,3, \\
        q=-3,\ldots,3\;.
    \end{array}
\end{equation}
Here, we transform a color image into a set of neighbor products (shift correlator) $V^k_{i,j}$ with $k=0,\ldots,48$, using the variance normalized image $\hat{I}$. The Gaussian filter used for local mean and variance is set to have kernel size $7\times 7$ and $\sigma=1.17$ based on similar values used in NIQE\cite{NIQE_2013}. The number of neighbor products is an additional parameter that we set to $7\times7$ that is a number larger than $3\times 3$ used in NIQE and BRISQUE and close to $64$ that is the number of features used in the discriminator architecture. The luminance channel is obtained using the BT.609 color matrix following the implementation of NIQE\cite{NIQE_2013}.

Figure \ref{fig:vnsc} shows the visual effect of the the VN+SC operation. A zoom on the tail of a zebra shows a visible correlation along the tail and less so in the stripes of the zebra. We use a VN+SC layer for each input of our discriminator, as shown in Figure \ref{fig:discr}. This reduces the amount of information received by the discriminator and, in particular, removes all color information. This follows closely the empirical findings of Ruderman\cite{Ruderman1994} to determine if an image looks natural. Color information will be considered by other terms of the loss function.

\textbf{Discriminator Architecture}. For the discriminator in adversarial training we use the system shown in Figure \ref{fig:discr}. Here, we use $4$--layer sequential CNNs with $3\times 3$ filters and stride $1$ except for the last layer that uses stride $2$ to downscale the features. The discriminator used with MGBPv1 in \cite{G-MGBP} had two minor differences: first, the CNNs shared parameters making the system recursive (imitating MGBPv1); and second, the progressive outputs of MGBPv1 entered the discriminator at different stages. Adversarial training with this discriminator resembles a Progressive GAN\cite{karras2018progressive} because it can adjust parameters to first solve the simpler problem of $2\times$ upscaling, and then follow with larger factors. But, at the same time, it is significantly different because a Progressive GAN system is neither multi--scale nor recursive. Overall, the major properties of our discriminator architecture, both for MGBPv1 and MGBPv2, are: input the RGB output of the generator in different resolutions (high resolution enter first, lower resolutions enter later); and, RGB images are converted into $49=7\times 7$ channels with the VN+SC modules. A 3D--convolutional version of the discriminator was used for video content in the AIM 2020 Video Extreme Super--Resolution Challenge \cite{fuoli2020aim} but did not succeed to achieve the same realistic effect observed in images. The temporal evolution of artificial details required for videos makes the problem significantly more difficult and currently unresolved. Therefore, here we will only consider the adversarial training strategy for image content.

\section{Learning}
\subsection{Configurations}
In the first version of MGBP (MGBPv1) the \emph{Analysis}, \emph{Synthesis}, \emph{Upscale}, \emph{Downscale} and \emph{CNN}--discriminator modules used $4$--layer dense networks\cite{huang2017densely} with a correspondent strided convolutional or transposed convolutional layer for downscale or upscale operations, respectively. We used $48$ features and growth rate $16$ within dense networks. For classic upscaler we configured a Bicubic and we set the upscaling filters as parameters to learn as proposed in \cite{MSLapSRN}.

In the second version, MGBPv2, the \emph{Analysis}, \emph{Synthesis}, \emph{Upscale}, \emph{Downscale} were configured with single convolutional layers. Here, we used different configurations depending on the upscaling factor. Table \ref{tab:mgbpv2_conf}--a shows the number of levels, the parameter $\mu$ and the number of convolutional layer features used for each upscaling factor. For small upscaling factors ($2$ and $3$) we chose a configuration almost equivalent to EDSR~\cite{Lim_2017_CVPR_Workshops} with only two levels and $32$ residual back--projection blocks.

For MGBP--3D the \emph{Analysis}, \emph{Synthesis}, \emph{Upscale}, \emph{Downscale} were configured using single 3D--convolutional layers as shown in Figure \ref{fig:mgbpv2}. Similar to MGBPv2, we used different configurations depending on the upscaling factors $4\times$ and $16\times$. Table \ref{tab:mgbpv2_conf}--b shows the number of levels, the parameter $\mu$ and the number of convolutional layer features used for each upscaling factor.

\begin{table}[t]
    \caption{MGBPv2 and MGBP--3D network configurations. High--fidelity performance is shown in Tables \ref{tab:fidelity} and \ref{tab:vxsr_track1}.}
    \label{tab:mgbpv2_conf}
    \centering
    \setlength{\tabcolsep}{4pt}
    a) MGBP--v2:
    \begin{tabular}{cccc}
        \hline
        Factor & $\mu$ & Levels & Channels per level\\ \hline
            2  &   32  &     2  & 192--128 \\
            3  &   32  &     2  & 192--128 \\
            4  &    4  &     4  & 192--128--64--32 \\
            8  &    2  &     5  & 192--128--64--32--16 \\
           16  &    2  &     6  & 256--192--128--92--48--9 \\
        \hline
    \end{tabular}

    b) MGBP--3D:
    \begin{tabular}{cccc}
               &       &        & \\
        \hline
        Factor & $\mu$ & Levels & Channels per level\\ \hline
            4  &    6  &     4  & 192--128--64--32 \\
           16  &    2  &     6  & 256--192--128--92--48--9 \\
        \hline
    \end{tabular}
\end{table}

For perceptual quality we concatenate a single noise channel, $W\cdot\mathcal{N}(0,1)$, to the bicubic upscaled input in MGBPv2 and the input at each resolution level in MGBPv1 as shown in Figures \ref{fig:mgbpv1} and \ref{fig:mgbpv2}. The amplitude of the noise, $W\in\mathbb{R}$, is set to zero for high--fidelity targets and its purpose is to help in the adversarial training for perceptual quality. The noise activates and deactivates the generation of artificial details. In MGBPv1 \cite{G-MGBP}, different i.i.d. noise channels are generated at each resolution. In MGBPv2 we generate a single noise channel at the highest resolution, that moves along with the input image to enter the network at different scales by using \emph{Analysis} modules with different strides and becomes more simple to manipulate.

We denote $Y_{W=0}$ and $Y_{W=1}$ the outputs of the generator architecture using noise amplitudes $W=0$ and $W=1$, respectively. In all the experiments and challenges that we present here, a high--resolution image has been resized using a bicubic downscaler. Let $S_f$ represent a bicubic downscaler that reduces the resolution by a factor $f\times f$ (horizontal and vertical). We would like our output $Y$ to be such that $S_f(X)=S_f(Y)$ where $X$ is the ground truth image. This is a prior knowledge that we will enforce in our loss functions during optimization.

\subsection{Learning High Fidelity}
\label{ssec:HF_loss}
We use $\mathcal{L}^{L1}(x,y)=\mathbb{E}\left[|x-y|\right]$ at several resolutions to define the total loss function:
\begin{align}
    \mathcal{L}(Y, X; \theta) = \; & \mathcal{L}^{L1}(Y_{W=0}, X) \; + \nonumber \\
                                   & \sum_{k\in\{2,4,\ldots, f\}}\mathcal{L}^{L1}(S_k(Y_{W=0}), S_k(X)) \;.
\end{align}
The first term is our main target, this is, to recover the high--resolution image. The sum over downscaling factors is meant to progressively give easier targets and enforce the downscaling model. After every epoch we evaluate the current model using $\mathcal{L}^{L2}(x,y)=\mathbb{E}\left[(x-y)^2\right]$ in the validation metric:
\begin{align}
    \mathcal{V}(Y; \theta) = \mathcal{L}^{L2}(Y_{W=0}, X) \;.
\end{align}
We recorded the best models according to this metric (directly related to PSNR) during the training process.

The input $X$ represents a 2D image or a 3D spatio--temporal cube for video content. In both MGBP--v2 and MGBP--3D the input $X$ correspond to the bicubic upscaled content with the same resolution as the output $Y$. For video content this means that the network learns to map a number of input frames to the same number of output frames. This is significantly different than the many--to--one frame mapping approach commonly used for video SR training and it is a consequence of the \emph{Cube--to--Cube} property of MGBP--3D.

\begin{figure}
    \centering
    \includegraphics[width=0.9\linewidth]{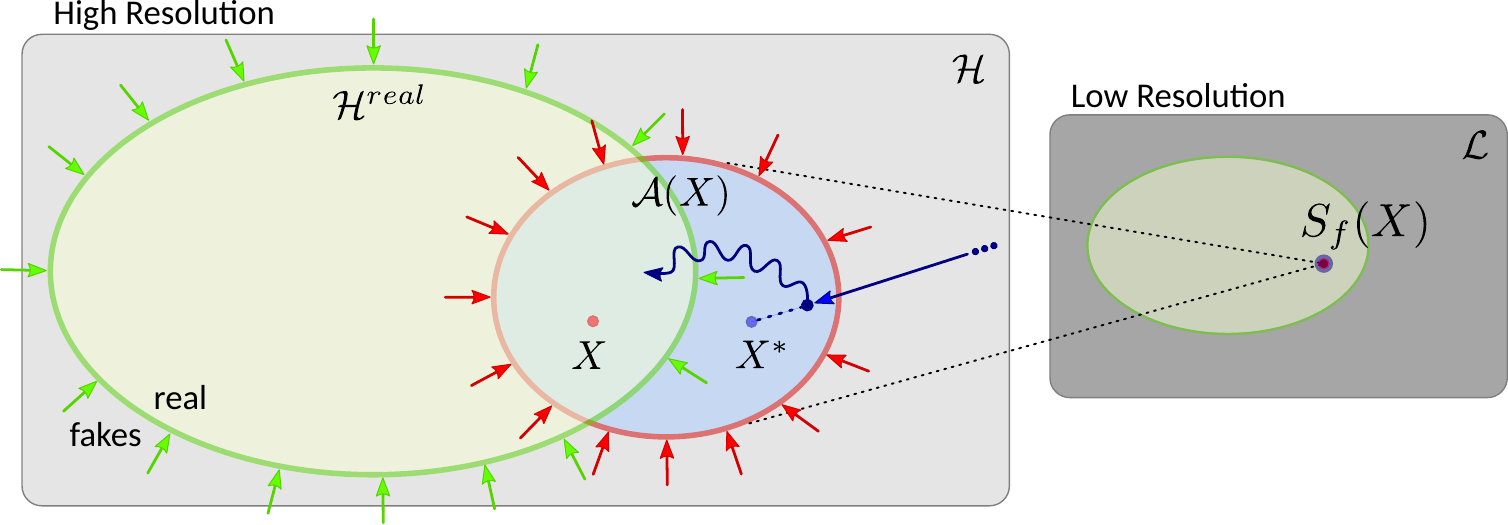}
    \caption{Our loss function tries to: look real by moving into $\mathcal{H}^{real}$ (GAN and CX loss), enforce a downscaling model by moving into $\mathcal{A}(X)$, and be reachable by latent space interpolation from the optimal distortion solution $X^*$ (distortion loss). \label{fig:strategy_losses}}
\end{figure}

\subsection{Learning Perceptual Quality}
For perceptual training we use a generative adversarial approach, alternating training steps for the generator and the discriminator, following the Relativistic GAN from \cite{jolicoeur2018relativistic}. For total loss we use the following expression:
\begin{align}
    \mathcal{L}(Y, X; \theta) = \; & 0.001 \cdot {\color{mygreen}\mathcal{L}^{RSGAN}_G(Y_{W=1})} \; + \nonumber \\
                                & 10 \cdot {\color{red}\tfrac{1}{|C|}\sum_{k\in C}\mathcal{L}^{L1}(S_k(Y_{W=1}), S_k(X))} \; + \nonumber \\
                                & 0.1 \cdot {\color{mygreen}\mathcal{L}^{CX}(Y_{W=1}, X)} \; + \nonumber \\
                                & 10 \cdot {\color{blue}\mathcal{L}^{L1}(Y_{W=0}, X)} \; + \nonumber \\
                                & 10 \cdot {\color{red}\tfrac{1}{|C|}\sum_{k\in C}\mathcal{L}^{L1}(S_k(Y_{W=0}), S_k(X))} \;, \label{eq:total_loss}
\end{align}
where the color is associated with the targets displayed in Figure \ref{fig:strategy_losses} and $C=\{2,4,\ldots,f\}$ is the set of upscaling factors. This is, green losses try to enter in the set of real images with $Y_{W=1}$; red losses try to enter the set of aliased images with both $Y_{W=1}$ and $Y_{W=0}$; and, blue losses try to recover the ground truth high--resolution image with $Y_{W=0}$.

The red losses in \eqref{eq:total_loss} enforcing the downscale model are also refer here as \emph{cycle} losses. This is because the output images are brought back to the input domain for comparison with the original input.

In particular, $\mathcal{L}^{CX}$ is the \emph{contextual loss} as defined in \cite{mechrez2018contextual} using features from \emph{conv3--4} of a VGG--19 network as suggested in \cite{mechrez2018Learning}. Ablation tests performed in the next section show the effectiveness of this loss function to improve perceptual quality while maintaining a reasonable level of distortion. Next, the Relativistic GAN loss follows the definition in \cite{jolicoeur2018relativistic}, given by:
\begin{align}
\mathcal{L}_D^{RSGAN} = & -\mathbb{E}_{(R,F)}\left[ \log (\text{sigmoid}(C(R)-C(F))) \right] \;,\nonumber \\
\mathcal{L}_G^{RSGAN} = & -\mathbb{E}_{(R,F)}\left[ \log (\text{sigmoid}(C(F)-C(R))) \right] \;.
\end{align}
Here, $C$ is the output of the discriminator before the sigmoid function, as shown in Figure \ref{fig:discr}. And $R$ and $F$ are the sets of real and fake inputs to the discriminator, given by:
\begin{align}
    F = & \left\{ Y_{W=1}, S_{2}(Y_{W=1}), S_{4}(Y_{W=1}), \ldots \right\} \;,  \nonumber \\
    R = & \left\{ X, S_{2}(X), S_{4}(X), \ldots \right\} \;.
\end{align}
After every epoch we evaluated the current model according to the validation metric based on the NIQE\cite{NIQE_2013} index:
\begin{align}
    \mathcal{V}(Y; \theta) = \mathbb{E}\Big[ & NIQE(Y_{W=1}) \Big] \;.
\end{align}
This metric works as a simple rule to help identify models that generate realistic images in the full resolution.

\begin{table}
\caption{Quantitative evaluation of MGBPv1 and MGBPv2 for image super--resolution with high--fidelity target.} \label{tab:fidelity}
\centering
\resizebox{\linewidth}{!}{
\begin{tabular}{lccccccccc}
    \hline
    & &
    \multicolumn{2}{c}{Set14} &
    \multicolumn{2}{c}{BSDS100} &
    \multicolumn{2}{c}{Urban100} &
    \multicolumn{2}{c}{Manga109} \\
    Algorithm & & PSNR & SSIM & PSNR & SSIM & PSNR & SSIM & PSNR & SSIM \\
    \hline
    Bicubic & \multirow{16}{*}{$2\times$}                          &  30.34  &  0.870  &  29.56  &  0.844  &  26.88  &  0.841  &  30.84  &  0.935 \\
    A+~\cite{Timofte_2014a}                                      & &  32.40  &  0.906  &  31.22  &  0.887  &  29.23  &  0.894  &  35.33  &  0.967 \\
    FSRCNN~\cite{Dong_2016a}                                     & &  32.73  &  0.909  &  31.51  &  0.891  &  29.87  &  0.901  &  36.62  &  0.971 \\
    SRCNN~\cite{Dong_2014a}                                      & &  32.29  &  0.903  &  31.36  &  0.888  &  29.52  &  0.895  &  35.72  &  0.968 \\
    MSLapSRN~\cite{MSLapSRN}                                     & &  33.28  &  0.915  &  32.05  &  0.898  &  31.15  &  0.919  &  37.78  &  0.976 \\
    VDSR~\cite{Kim_2016_VDSR}                                    & &  32.97  &  0.913  &  31.90  &  0.896  &  30.77  &  0.914  &  37.16  &  0.974 \\
    LapSRN~\cite{LapSRN}                                         & &  33.08  &  0.913  &  31.80  &  0.895  &  30.41  &  0.910  &  37.27  &  0.974 \\
    DRCN~\cite{Kim_2016_DRCN}                                    & &  32.98  &  0.913  &  31.85  &  0.894  &  30.76  &  0.913  &  37.57  &  0.973 \\
    D-DBPN~\cite{DBPN2018}                                       & &  33.85  &  0.919  &  32.27  &  0.900  &  32.70  &  0.931  &  39.10  &  0.978 \\
    EDSR~\cite{Lim_2017_CVPR_Workshops}                          & &  33.92  &  0.919  &  32.32  &  0.901  &  32.93  &  0.935  &  39.10  &  0.977 \\
    RDN~\cite{zhang2020residual}                                 & & \textbf{34.28} & \textbf{0.924} & \textbf{32.46} & \textbf{0.903} & \textbf{33.36} & \textbf{0.939} & \textbf{39.74} & \textbf{0.979} \\
    RCAN~\cite{zhang2018rcan}                                    & &  34.12  &  0.921  &  32.41  & \textbf{0.903} &  33.34  &  0.938  &  39.44  & \textbf{0.979} \\
    \rowcolor{lightgray} MGBPv1~\cite{PNavarrete_2019a}          & &  33.27  &  0.915  &  31.99  &  0.897  &  31.37  &  0.920  &  37.92  &  0.976 \\
    \rowcolor{lightgray} MGBPv2                                  & &  34.18  &  0.922  &  31.70  &  0.887  &  33.31  &  0.936  &  39.42  &  0.978  \\
    \noalign{\smallskip}
    \hline
    \noalign{\smallskip}
    Bicubic & \multirow{8}{*}{$3\times$}                           &  27.55  &  0.774  &  27.21  &  0.739  &  24.46  &  0.735  &  26.95  &  0.856 \\
    SRCNN~\cite{Dong_2014a}                                      & &  29.30  &  0.822  &  28.41  &  0.786  &  26.24  &  0.799  &  30.48  &  0.912 \\
    MSLapSRN~\cite{MSLapSRN}                                     & &  29.97  &  0.836  &  28.93  &  0.800  &  27.47  &  0.837  &  32.68  &  0.939 \\
    LapSRN~\cite{LapSRN}                                         & &  29.87  &  0.832  &  28.82  &  0.798  &  27.07  &  0.828  &  32.21  &  0.935 \\
    EDSR~\cite{Lim_2017_CVPR_Workshops}                          & &  30.52  &  0.846  &  29.25  &  0.809  &  28.80  &  0.865  &  34.17  &  0.948 \\
    RDN~\cite{zhang2020residual}                                 & &  \textbf{30.74}  & \textbf{0.850} & \textbf{29.38} & \textbf{0.812} &  \textbf{29.18}  &  \textbf{0.872}  & \textbf{34.81} & \textbf{0.951} \\
    \rowcolor{lightgray} MGBPv2                                  & &  30.60  &  0.845  &  29.10  &  0.803  &  29.15  &  0.866  &  34.31  &  0.947  \\
    \noalign{\smallskip}
    \hline
    \noalign{\smallskip}
    Bicubic & \multirow{16}{*}{$4\times$}                          &  26.10  &  0.704  &  25.96  &  0.669  &  23.15  &  0.659  &  24.92  &  0.789 \\
    A+~\cite{Timofte_2014a}                                      & &  27.43  &  0.752  &  26.82  &  0.710  &  24.34  &  0.720  &  27.02  &  0.850 \\
    FSRCNN~\cite{Dong_2016a}                                     & &  27.70  &  0.756  &  26.97  &  0.714  &  24.61  &  0.727  &  27.89  &  0.859 \\
    SRCNN~\cite{Dong_2014a}                                      & &  27.61  &  0.754  &  26.91  &  0.712  &  24.53  &  0.724  &  27.66  &  0.858 \\
    MSLapSRN~\cite{MSLapSRN}                                     & &  28.26  &  0.774  &  27.43  &  0.731  &  25.51  &  0.768  &  29.54  &  0.897 \\
    VDSR~\cite{Kim_2016_VDSR}                                    & &  28.03  &  0.770  &  27.29  &  0.726  &  25.18  &  0.753  &  28.82  &  0.886 \\
    LapSRN~\cite{LapSRN}                                         & &  28.19  &  0.772  &  27.32  &  0.728  &  25.21  &  0.756  &  29.09  &  0.890 \\
    DRCN~\cite{Kim_2016_DRCN}                                    & &  28.04  &  0.770  &  27.24  &  0.724  &  25.14  &  0.752  &  28.97  &  0.886 \\
    D-DBPN~\cite{DBPN2018}                                       & &  28.82  &  0.786  &  27.72  &  0.740  &  26.54  &  0.795  &  31.18  &  0.914 \\
    EDSR~\cite{Lim_2017_CVPR_Workshops}                          & &  28.80  &  0.788  &  27.71  &  0.742  &  26.64  &  0.803  &  31.02  &  0.915 \\
    RDN~\cite{zhang2020residual}                                 & &  \textbf{29.01}  & \textbf{0.791} &  \textbf{27.85}  & \textbf{0.745} &  27.01  &  0.812  & \textbf{31.74} & \textbf{0.921} \\
    RCAN~\cite{zhang2018rcan}                                    & &  28.87  &  0.789  &  27.77  &  0.744  &  26.82  &  0.809  &  31.22  &  0.917 \\
    \rowcolor{lightgray} MGBPv1~\cite{PNavarrete_2019a}          & &  28.43  &  0.778  &  27.42  &  0.732  &  25.70  &  0.774  &  30.07  &  0.904 \\
    \rowcolor{lightgray} MGBPv2                                  & &  29.00  &  0.790  &  27.87  &  \textbf{0.745}  &  \textbf{27.08}  &  \textbf{0.820}  &  31.45  &  0.917  \\
    \noalign{\smallskip}
    \hline
    \noalign{\smallskip}
    Bicubic & \multirow{15}{*}{$8\times$}                          &  23.19  &  0.568  &  23.67  &  0.547  &  20.74  &  0.516  &  21.47  &  0.647 \\
    A+~\cite{Timofte_2014a}                                      & &  23.98  &  0.597  &  24.20  &  0.568  &  21.37  &  0.545  &  22.39  &  0.680 \\
    FSRCNN~\cite{Dong_2016a}                                     & &  23.93  &  0.592  &  24.21  &  0.567  &  21.32  &  0.537  &  22.39  &  0.672 \\
    SRCNN~\cite{Dong_2014a}                                      & &  23.85  &  0.593  &  24.13  &  0.565  &  21.29  &  0.543  &  22.37  &  0.682 \\
    MSLapSRN~\cite{MSLapSRN}                                     & &  24.57  &  0.629  &  24.65  &  0.592  &  22.06  &  0.598  &  23.90  &  0.759 \\
    VDSR~\cite{Kim_2016_VDSR}                                    & &  24.21  &  0.609  &  24.37  &  0.576  &  21.54  &  0.560  &  22.83  &  0.707 \\
    LapSRN~\cite{LapSRN}                                         & &  24.44  &  0.623  &  24.54  &  0.586  &  21.81  &  0.582  &  23.39  &  0.735 \\
    D-DBPN~\cite{DBPN2018}                                       & &  25.13  &  0.648  &  24.88  &  0.601  &  22.83  &  0.622  &  25.30  &  0.799 \\
    EDSR~\cite{Lim_2017_CVPR_Workshops}                          & &  24.94  &  0.640  &  24.80  &  0.596  &  22.47  &  0.620  &  24.58  &  0.778 \\
    RDN~\cite{zhang2020residual}                                 & &  \textbf{25.38}  &  \textbf{0.654}  &  25.01  &  \textbf{0.606}  &  \textbf{23.04}  &  \textbf{0.644}  &  \textbf{25.48}  &  \textbf{0.806} \\
    RCAN~\cite{zhang2018rcan}                                    & &  25.23  &  0.651  &  24.98  &  \textbf{0.606}  &  23.00  &  0.645  &  25.24  &  0.803 \\
    \rowcolor{lightgray} MGBPv1~\cite{PNavarrete_2019a}          & &  24.82  &  0.635  &  24.67  &  0.592  &  22.21  &  0.603  &  24.12  &  0.765 \\
    \rowcolor{lightgray} MGBPv2                                  & &  25.37  &  0.652  &  \textbf{25.08}  &  \textbf{0.606}  &  22.99  &  0.640  &  25.07  &  0.795 \\
    \hline
    \end{tabular}
}
\end{table}

\section{Performance}

\begin{figure}
    \centering
    \includegraphics[width=0.9\linewidth]{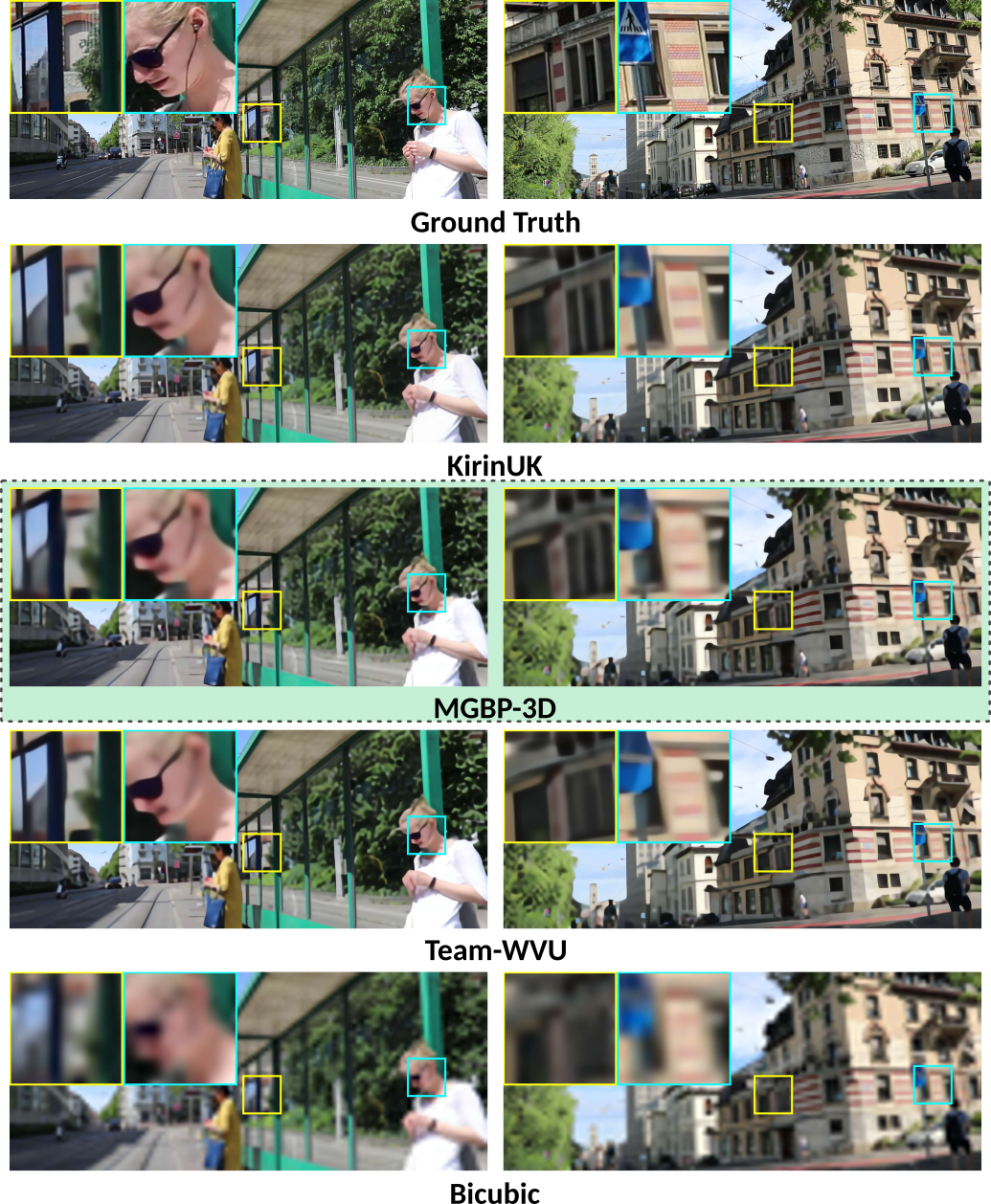}
    \caption{Qualitative evaluation of $16\times$ video SR results from the AIM 2020 Video Extreme Super--Resolution Challenge \cite{fuoli2020aim}. MGBP--3D frames can be downloaded from \href{https://www.dropbox.com/s/few4dpe7g68mc60/BOE_Perceptual_SmoothMotion_AllFrames.zip}{here.} \label{fig:vxsr_track1}}
\end{figure}

\begin{table*}
    \caption{Quantitative comparison on REDS4 for $4\times$ video Super--Resolution. MGBP--3D frames can be downloaded from \href{https://www.dropbox.com/s/pnheyf8bv6s6xqe/REDS4_validation_output.zip}{here.}}
    \label{tab:reds4}
    \centering
    \begin{tabular}{lcccccccc|cc}
        \hline
        Method & \multicolumn{2}{c}{Clip\_000} & \multicolumn{2}{c}{Clip\_011} & \multicolumn{2}{c}{Clip\_015} & \multicolumn{2}{c}{Clip\_020} & \multicolumn{2}{c}{Average} \\
                                      &  PSNR &   SSIM &  PSNR &   SSIM &  PSNR &   SSIM &  PSNR &   SSIM &  PSNR &   SSIM \\ \hline
        Bicubic                       & 24.55 & 0.6489 & 26.06 & 0.7261 & 28.52 & 0.8034 & 25.41 & 0.7386 & 26.14 & 0.7292 \\
        RCAN~\cite{zhang2018image}    & 26.17 & 0.7371 & 29.34 & 0.8255 & 31.85 & 0.8881 & 27.74 & 0.8293 & 28.78 & 0.8200 \\
        TOFlow~\cite{xue2019video}    & 26.52 & 0.7540 & 27.80 & 0.7858 & 30.67 & 0.8609 & 26.92 & 0.7953 & 27.98 & 0.7990 \\
        DUF~\cite{jo2018deep}         & 27.30 & 0.7937 & 28.38 & 0.8056 & 31.55 & 0.8846 & 27.30 & 0.8164 & 28.63 & 0.8251 \\
        EDVR~\cite{wang2019edvr}      & 28.01 & 0.8250 & 32.17 & 0.8864 & 34.06 & 0.9206 & 30.09 & 0.8881 & 31.09 & 0.8800 \\
        \rowcolor{lightgray} MGBP--3D & 27.04 & 0.7793 & 29.11 & 0.8210 & 31.83 & 0.8855 & 27.77 & 0.8286 & 28.94 & 0.8286 \\ \hline
      \end{tabular}
\end{table*}

\subsection{Fidelity Evaluation}
In image SR we use DIV2K\cite{Agustsson_2017_CVPR_Workshops} and FLICKR--2K datasets for training and the following datasets for test: Set--14\cite{zeyde2010single}, BSDS--100\cite{martin2001database}, Urban--100\cite{huang2015single} and Manga--109\cite{matsui2017sketch-based}. Impaired images were obtained by downscaling and then upscaling ground truth images, using bicubic scaler, with scaling factors: $2\times$, $3\times$, $4\times$ and $8\times$. For evaluation we measure PSNR and SSIM on the Y--channel using the Matlab code from \cite{psnr-ssim-y}.

We follow the training settings from \cite{Lim_2017_CVPR_Workshops}. In each training batch, we randomly take $16$ impaired patches from our training set ($800$ DIV2K plus $2,650$ FLICKR--2K images). We consider both MGBPv1 and MGBPv2 models for each upscaling factor $f=2, 4$ and $8$. We use patch size $48f\times 48f$, for $f=2, 3, 4$ and $8$. We augment the patches by random horizontal/vertical flipping and rotating $90^\circ$. We use Adam optimizer\cite{kingma2014adam} with learning rate initialized to $10^{-4}$ and decreased by half every $200,000$ back--propagation steps.

Table \ref{tab:fidelity} shows the performance at different upscaling factors. We observe that MGBPv2 performs very close to the best benchmarks, and performs better for large upscale factors $4\times$ and $8\times$. MGBPv1 performance is lower than MGBPv2, but it is only outperformed by systems with large number of parameters like EDSR, DBPN, RDN, RCAN and MGBPv2. At small upscaling factors $2\times$ and $3\times$ we used an MGBPv2 configuration very similar to EDSR, with $32$ residual blocks in two levels, but MGBPv2 clearly outperforms EDSR. This indicate the advantage of the residual back--projection block compared to conventional residual blocks. The multigrid recursion is effective in improving performance for large upscaling factors. It must be noted that the major trend in CNN architectures is to do all processing at low resolution and use PixelShuffle layers to quickly move to the highest resolution. MGBP balances this process by doing most work on the lowest resolutions and decrease the amount of work exponentially as it goes to higher resolutions.

In video SR we trained MGBP--3D using the REDS dataset \cite{Nah_2019_CVPR_Workshops_REDS} for $4\times$ and Vid3oC \cite{kim2019vid3oc} for $16\times$ upscaling. In particular, for $4\times$ we used the REDS4 configuration as specified in \cite{wang2019edvr}, training MGBP--3D with $266$ video sequences and testing with $4$ video sequences. Quantitative results are included in Table \ref{tab:reds4}. Here, MGBP--3D outperforms the image SR network RCAN~\cite{zhang2018image} based on attention modules, the video SR network TOFlow~\cite{xue2019video} using optical flow estimation, and the video SR network DUF~\cite{jo2018deep} using dynamic upscaling filters. But MGBP--3D is still far from the video SR network EDVR~\cite{wang2019edvr} using deformable convolutions. In extreme video SR ($16\times$) we refer to the results of MGBP--3D in the AIM 2020 Challenge on Video Extreme Super--Resolution~\cite{fuoli2020aim} with quantitative evaluations shown in Table \ref{tab:vxsr_track1} and qualitative evaluations shown in Figure \ref{fig:vxsr_track1}. Team KirinUK won the competition introducing EVESR--Net, using deformable convolutions for alignment and new attention modules \cite{fuoli2020aim}. MGBP--3D shared the $2^{nd}$ place with Team--WVU that also used an architecture based on deformable convolutions. Overall, the result for both $4\times$ and $16\times$ indicate that MGBP--3D gains advantage compared to other methods for large upscale factors. The arrangement of the MGBP--3D architecture due to the multigrid strategy is particularly effective considering that the architecture only uses 3D--convolutions and ReLU units. This is in contrast to other methods using attention, deformable convolutions, warping or other non–linear modules.

\begin{table}
    \caption{Quantitative results for AIM 2020 Video Extreme Super--Resolution Challenge ($16\times$)\cite{fuoli2020aim}.} \label{tab:vxsr_track1}
    \centering
    \begin{tabular}{clrrrc} \hline
              Fidelity & \multirow{2}{*}{Team} & \multirow{2}{*}{PSNR} & \multirow{2}{*}{SSIM} & \multirow{2}{*}{Runtime} & Perceptual \\
                  Rank &            &                &                 &        & Rank \\ \hline
                     1 & KirinUK    & \textbf{22.83} & \textbf{0.6450} & 6.1s   & 1 \\
\rowcolor{lightgray} 2 & MGBP--3D   & 22.48          & 0.6304          & 4.83s  & 4 \\
                     2 & Team-WVU   & 22.48          & 0.6378          & 4.90s  & 2 \\
                     3 & sr\_xxx    & 22.43          & 0.6353          & 4s     & 5 \\
                     4 & ZZX        & 22.28          & 0.6321          & 4s     & 3 \\
                     5 & lyl        & 22.08          & 0.6256          & 13s    & 6 \\
                     6 & TTI        & 21.91          & 0.6165          & 0.249s & -- \\
                     7 & CET\_CVLab & 21.77          & 0.6112          & 0.04s  & 7 \\
                       & Bicubic    & 20.69          &  0.5770         &        & \\ \hline
    \end{tabular}
\end{table}

\subsection{Perceptual Evaluation}
For image perceptual quality evaluations we refer to the results of MGBPv1 and MGBPv2 in the first challenge on perceptual image SR, PIRM--SR 2018~\cite{blau20182018}, and the first extreme SR challenge, AIM Extreme--SR 2019~\cite{lugmayr2019aim}, respectively.

\begin{figure}
    \centering
    \includegraphics[width=0.9\linewidth]{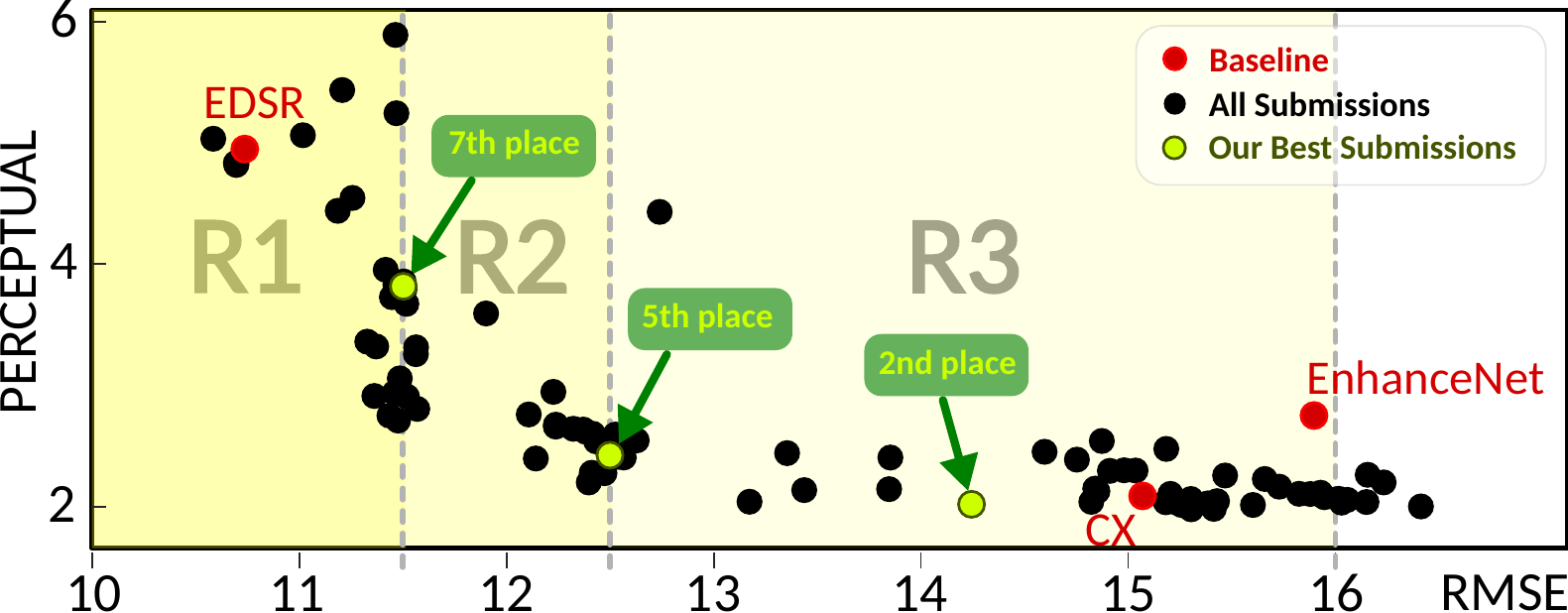}
    \caption{Perception--distortion plane with average scores in the test set showing all submissions, from all teams in PIRM--SR 2018\cite{blau20182018}. Our best scores are shown in green color together with the final ranking in PIRM--SR Challenge 2018. The perception--distortion trade--off is clearly visible with an empty area in the lower--left corner.\label{fig:pirm2018_final_results}}
\end{figure}

Table \ref{tab:pirm_results} shows our best average scores and rankings in the PIRM--SR Challenge 2018\cite{blau20182018} for Region 1 ($RMSE\leqslant 11.5$), Region 2 ($11.5<RMSE\leqslant 12.5$) and Region 3 ($12.5<RMSE\leqslant 16$). Figure \ref{fig:pirm2018_final_results} shows all submissions in the perceptual--distortion plane, including the baseline methods: EDSR\cite{Lim_2017_CVPR_Workshops}, CX\cite{mechrez2018contextual} and EnhanceNet\cite{Sajjadi_2017_ICCV}.

\begin{table*}[t]
    \scriptsize
    \caption{PIRM 2018 Challenge results. MOS tests were performed only for top submissions~\cite{blau20182018}.}
    \centering
    \begin{tabular}{@{}llccccllccccllccc@{}}
        \multicolumn{5}{c}{\textbf{Region 1}} & \phantom{abc}& \multicolumn{5}{c}{\textbf{Region 2}} & \phantom{abc} & \multicolumn{5}{c}{\textbf{Region 3}}\\
        \cmidrule{1 - 5} \cmidrule{7 - 11} \cmidrule{13 - 17} Rank & Team & \makecell{PI} & RMSE & MOS & & Rank & Team & \makecell{PI} & RMSE & MOS & & Rank &Team & \makecell{PI} & RMSE & MOS \\
        \cmidrule{1 - 5} \cmidrule{7 - 11} \cmidrule{13 - 17}
        $1$ & IPCV \cite{vasu2018analyzing} & 2.709 & 11.48 & 2.22 && $1$ & TTI \cite{DBPN2019} & 2.199 & 12.40 & 2.17 && $1$ & SuperSR \cite{wang2018esrgan} & 1.978 & 15.30 & 2.64 \\
        $2$ & MCML \cite{cheon2018generative} & 2.750 & 11.44 & 1.87 && $2$ & IPCV \cite{vasu2018analyzing} & 2.275 & 12.47 & 2.43 && \cellcolor{lightgray}$2$ & \cellcolor{lightgray}MGBPv1 \cite{G-MGBP} & \cellcolor{lightgray}2.019 & \cellcolor{lightgray}14.24 & \cellcolor{lightgray}2.61 \\
        $3$ & SuperSR \cite{wang2018esrgan} & 2.933 & 11.50 & 2.19 && $2$ & MCML \cite{choi2018deep} & 2.279 & 12.41 & 2.47 && $3$ & IPCV \cite{vasu2018analyzing} & 2.013 & 15.26 & 2.60 \\
        $3$ & TTI \cite{DBPN2019} & 2.938 & 11.46 & 1.88 && $4$ & SuperSR \cite{wang2018esrgan} & 2.424 & 12.50 &-- && $4$ & AIM \cite{vu2018perception} & 2.013 & 15.60 & -- \\
        $5$ & AIM \cite{vu2018perception} & 3.321 & 11.37 & -- && \cellcolor{lightgray}$5$ & \cellcolor{lightgray}MGBPv1 \cite{G-MGBP} & \cellcolor{lightgray}2.484 & \cellcolor{lightgray}12.50 &\cellcolor{lightgray} -- && $5$ & TTI \cite{DBPN2019} & 2.040 & 13.17 & -- \\
        $6$ & DSP--whu & 3.728 & 11.45 & -- && $6$ & AIM \cite{vu2018perception} & 2.600 & 12.42 & -- && $6$ & Haiyun \cite{luo2018bi} & 2.077 & 15.95 & -- \\
        \cellcolor{lightgray}$7$ & \cellcolor{lightgray}MGBPv1 \cite{G-MGBP} & \cellcolor{lightgray}3.817 & \cellcolor{lightgray}11.50 &\cellcolor{lightgray} -- && $7$ & REC--SR \cite{kuldeep2018scale} & 2.635 & 12.37 & -- && $7$ & gayNet & 2.104 & 15.88 & -- \\
        $7$ & REC--SR \cite{kuldeep2018scale} & 3.831 & 11.46 & -- && $8$ & DSP--whu & 2.660 & 12.24 & -- && $8$ & DSP--whu & 2.114 & 15.93 & -- \\
        $9$ & Haiyun \cite{luo2018bi} & 4.440 & 11.19 & -- && $9$ & XYN & 2.946 & 12.23 & -- && $9$ & MCML & 2.136 & 13.44 & -- \\
    \end{tabular}
    \label{tab:pirm_results}
\end{table*}

\begin{table}[t]
    \caption{AIM 2019 Extreme SR Challenge Track 2 Perceptual results and final rankings on the DIV8K test set ~\cite{lugmayr2019aim}.}
    \label{tab:aim_results}
    \centering
    \setlength{\tabcolsep}{4pt}
    \begin{tabular}{clccl}
        \hline
        Rank & Team/Method  & PSNR  & SSIM   & Runtime \\ \hline
        \rowcolor{lightgray} $1$  & MGBPv2 & 25.44 & 0.6551 & 47.11s \\
                             $2$  & TTI    & 25.26 & 0.6523 & 80s \\
                             $3$  & SRSTAR & 26.72 & 0.7285 & 40s \\
                             $4$  & SSRR   & 26.53 & 0.7246 & 40s \\
        \hline
    \end{tabular}
\end{table}

Compared to other submissions, we observe in Figure \ref{fig:pirm2018_final_results} that our system performs better in Region 3. Here, we achieve the $2^{nd}$ place within very small differences in perceptual scores but with significantly lower distortion. This shows the advantage of our training strategy to optimize the perception--distortion trade--off. In Regions 1 and 2 we were one among only two teams that reached the exact distortion limit ($11.5$ in Region 1 and $12.5$ in Region 2). We were able to achieve this by controlling the noise amplitude, without retraining the system. Our ranking lowers as the distortion target gets more difficult. We believe that this is caused by the small size of our system that becomes more important for low distortion targets, since we use only $281k$ parameters compared to $43M$ of the EDSR baseline in Region 1.

\begin{figure}
    \centering
    \includegraphics[width=0.92\linewidth]{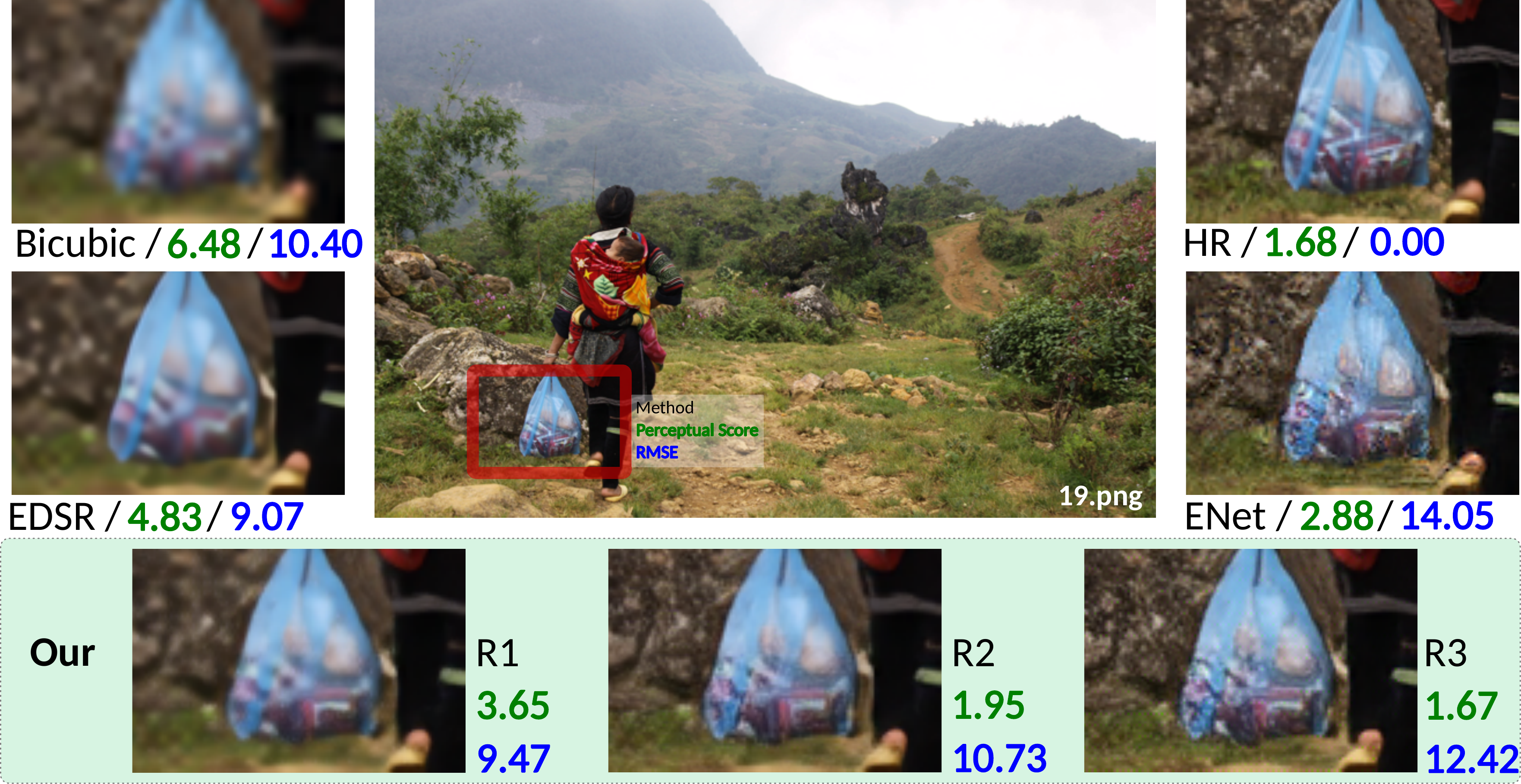} \medskip
    \hrule \medskip
    \includegraphics[width=0.92\linewidth]{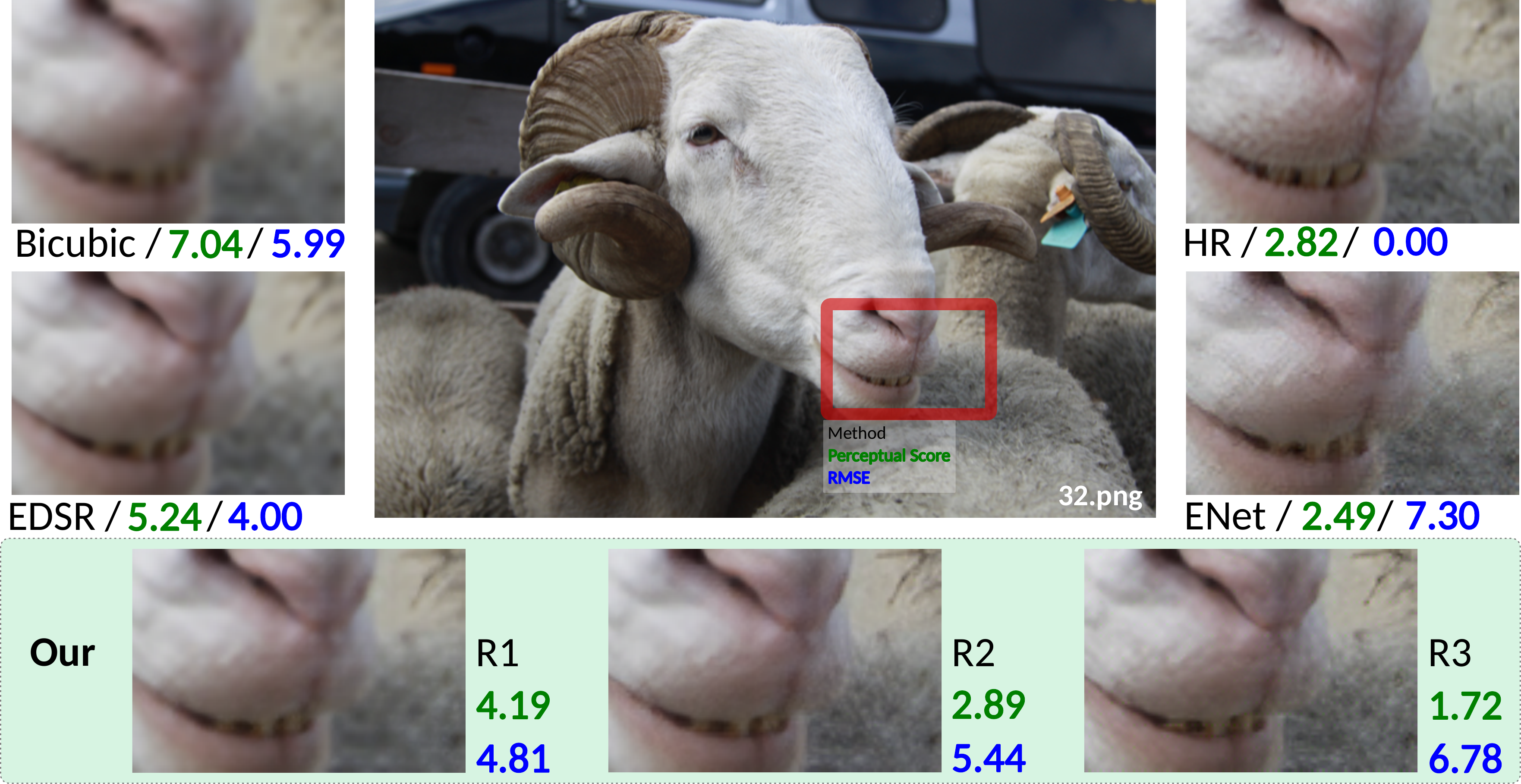} \medskip
    \hrule \medskip
    \includegraphics[width=0.92\linewidth]{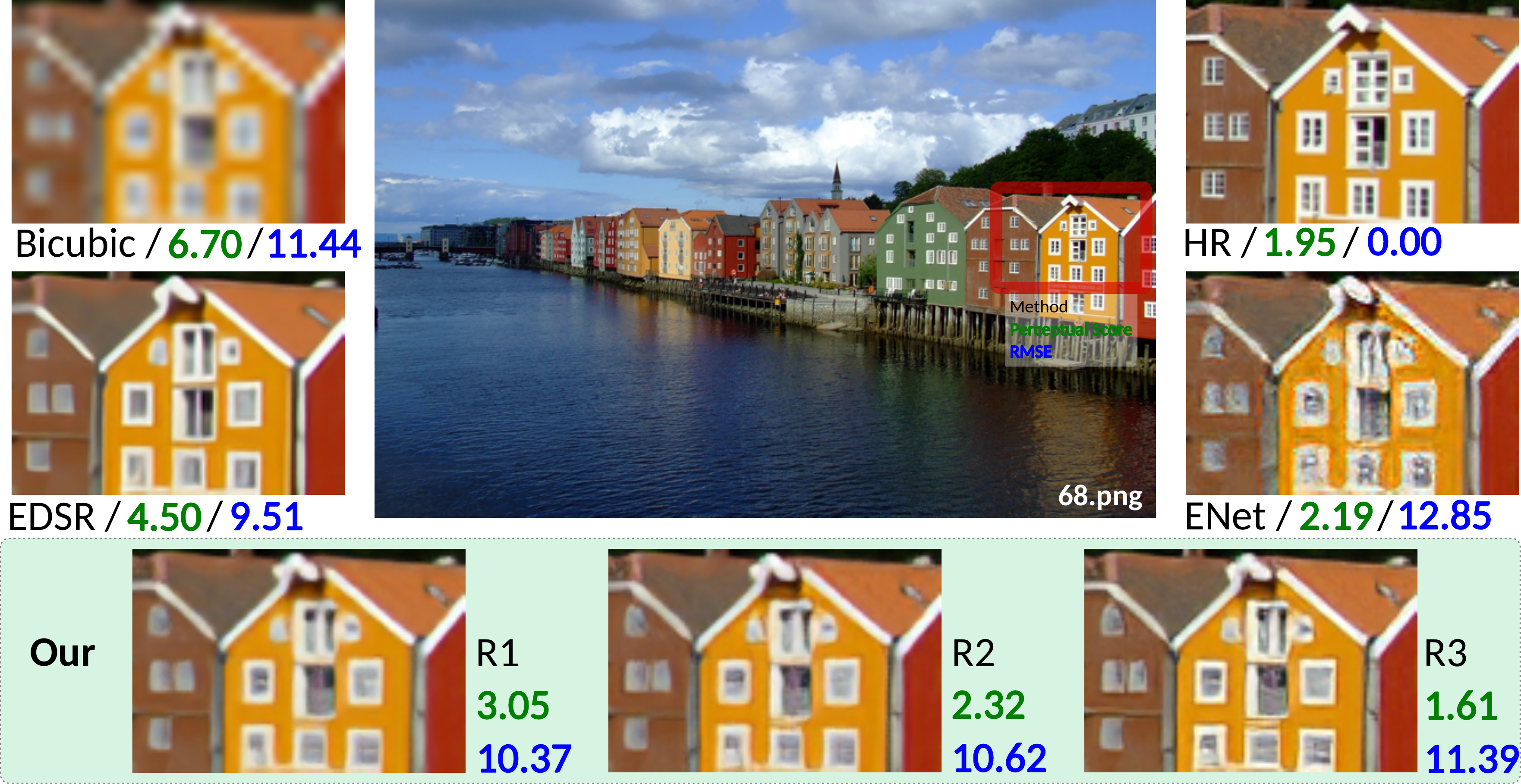}
    \caption{PIRM-SR 2018 comparisons of $4\times$ upscaling between our solutions in R1, R2 and R3 (see Figure \ref{fig:pirm2018_final_results}) and baseline methods in our validation set. Perceptual and distortion scores of whole images are shown in green and blue colors, respectively. \label{fig:pirm_comparison}}
\end{figure}

Figure \ref{fig:pirm_comparison} shows comparisons of our results with the baselines, using images from our validation set. We observe that in Region 3 we achieve better perceptual scores even compared to the original HR images. While we subjectively confirm this in some patches, we do not make the same conclusion after observing the whole images. Somehow, we believe that our design for adversarial training and validation strategy managed to overfit the perceptual scores. Nevertheless, we observe clear advantages to the baselines, showing better structure in textures and more consistent geometry in edges and shapes.

\begin{figure*}[t]
  \centering
  \includegraphics[width=.9\linewidth]{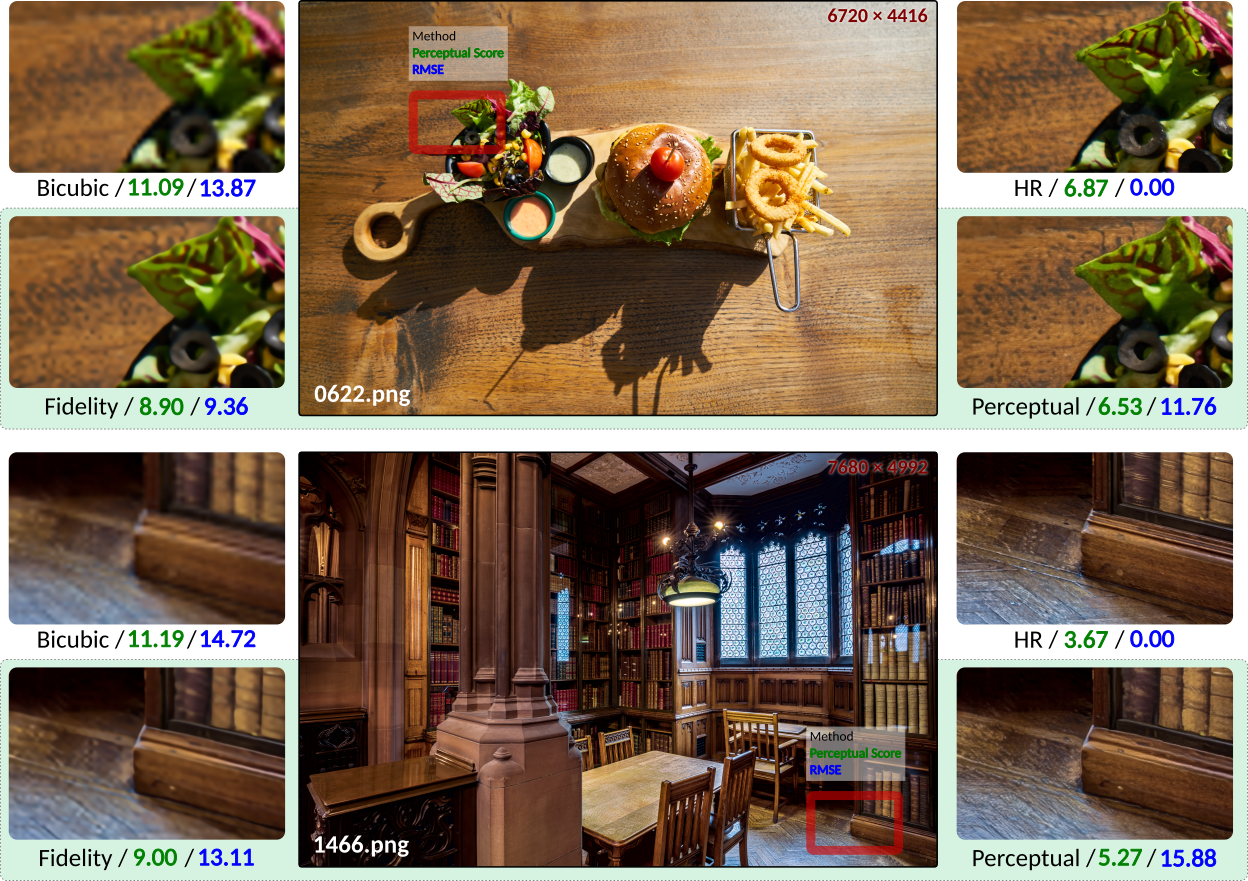}
  \caption{Example outputs and performance metrics of our MGBPv2 $16\times$ model used for AIM Extreme--SR 2019~\cite{lugmayr2019aim}. These images were selected from the DIV8K training set and we used them for validation purposes. \label{fig:aim_comparison2}}
\end{figure*}

In the Perceptual track of the Extreme--SR AIM 2019 challenge our results using MGBPv2 obtained the $1^{st}$ place in terms of a subjective ranking based on mean opinion scores (MOS) to measure perceptual quality. Our results obtained an average PSNR of $25.44$ dB in the full output images of the test set. This is, $0.18$ dB above the $2^{nd}$ place and $1.35$ dB below the best PSNR value in the Fidelity track.

Figure \ref{fig:aim_comparison2} shows two examples of our best results for the Fidelity and Perceptual tracks on images used for validation during our training process. These images show the values of RMSE (measuring fidelity) as well as the Perceptual index proposed in \cite{blau20182018} to objectively measure perceptual quality. Overall, the results are consistent with the perception/distortion trade--off in \cite{Blau_2018_CVPR}. For the image \texttt{622} we observe that our results for the Perceptual track achieve a Perceptual index better than original image. We attribute this result to a blurred background in the original image, that our system shows more focused and with sharper features. For image \texttt{1466}, the Perceptual index is clearly below those of the original image. According to our subjective evaluation, we observe clear differences in the fine level features like textbook spines. From a far away look these details become less perceptible, indicating that the Perceptual index correlates better with a close distance observer. This is probably caused by the resolution of example images used to adjust the Perceptual index, that are much smaller than 8K.

\begin{figure*}
  \centering
  \includegraphics[width=\linewidth]{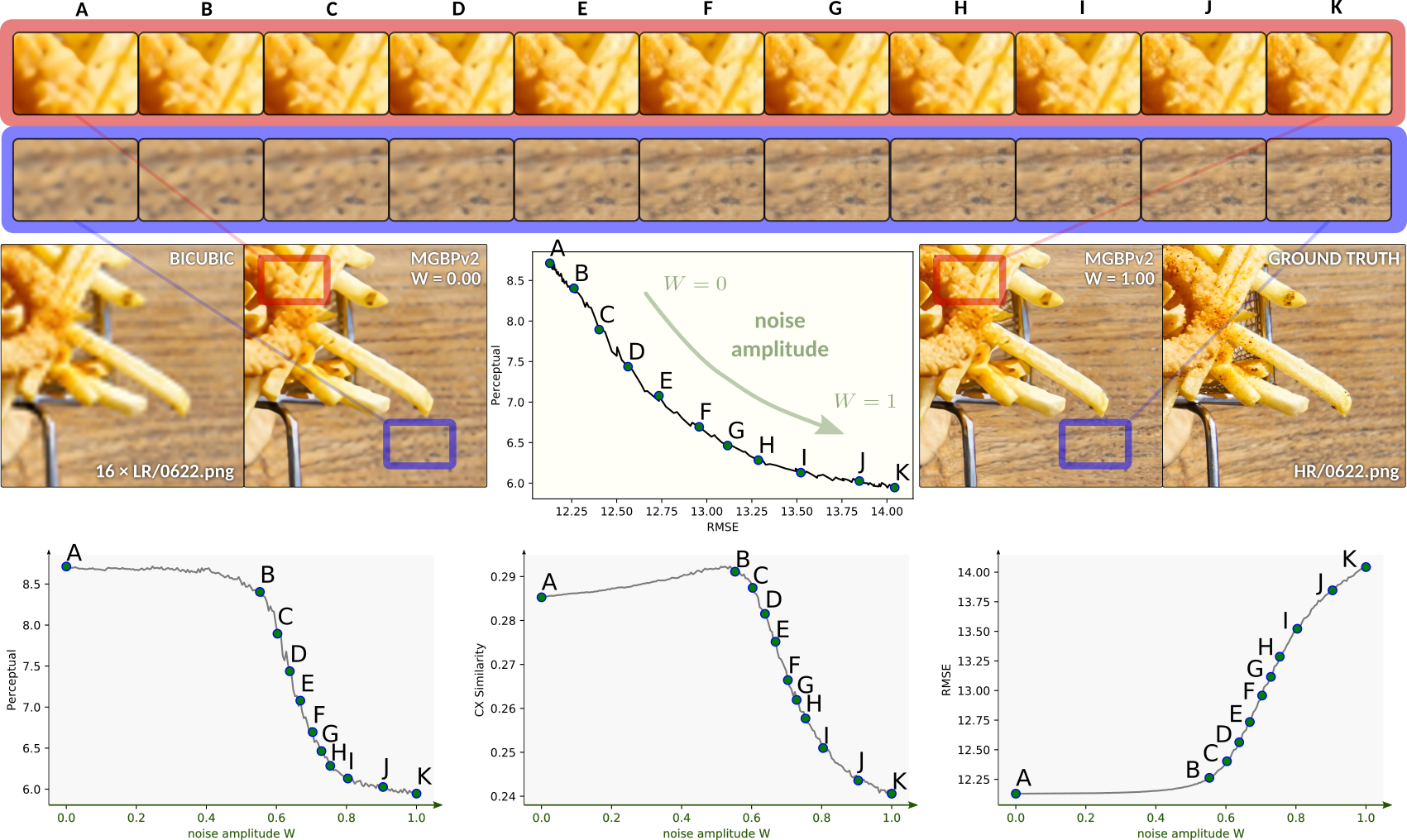}
  \caption{Evolution of perceptual and fidelity metrics when moving the input noise amplitude from $W=0$ to $W=1$ in our MGBPv2 $16\times$ system used for AIM Extreme--SR 2019~\cite{lugmayr2019aim}. \label{fig:trajectory2}}
\end{figure*}

\section{Analysis}
\subsection{Perception--Distortion Trajectory}
An essential part of our generative SR architecture is the noise input. The training strategy introduced in Section \ref{sec:strategy} teaches the system to optimize distortion when the noise is set to zero, and maximize perceptual quality when the noise is enabled. Thus, noise provides the randomness needed for natural images and represents the \emph{innovation jump} according to Figure \ref{fig:strategy_sets}.

After training, we are free to control the noise inputs. In particular, we can move the noise amplitude smoothly between $W=0$ and $W=1$ to inspect the path to jump from distortion to perception optimization. Figure \ref{fig:trajectory2} shows an example of this transition. Our training strategy does not optimize the trajectory in the perception--distortion plane, but only the corner cases of best distortion ($W=0$) and best perception ($W=1$). The corner cases are clearly verified in Figure \ref{fig:trajectory2}. At this point, it is unknown which trajectory will the the network take to move from one case to the other.

It is interesting to see in Figure \ref{fig:trajectory2} that the transition from best perception to best distortion happens within a narrow margin of $\Delta W = 0.2$ amplitude values. We also observe that the parametric curve in the perception--distortion plane looks like a monotonically non--increasing and convex function, similar to the optimal solution studied in \cite{Blau_2018_CVPR}. But, it is important to emphasize that the curve in Figure \ref{fig:trajectory2} is not optimal as we are not enforcing optimality. The results in Table \ref{tab:pirm_results} and Figure \ref{fig:pirm2018_final_results} clearly show that our perception--distortion curve, despite being concave, gets farther from optimality in low distortion regions.

Regarding image quality metrics, we see with no surprise that the \emph{Perceptual} index proposed for the PIRM--SR Challenge\cite{blau20182018} improves as noise increases, while the distortion measured by RMSE increases. We observed very similar results for the perceptual metrics NIQE and Ma, as well as the L1 distortion metric. More interesting is the transition observed in the \emph{contextual similarity} index. First, it behaves as a perceptual score with the CX similarity improving consistently as noise increases. Then, when the \emph{Perceptual} score seems to stall, but RMSE keeps increasing, the CX similarity changes to a distortion metric pattern, reducing as noise increases. This is consistent with the design target of \emph{CX similarity} to focus more on perceptual quality while maintaining a reasonable level of distortion\cite{mechrez2018contextual}.

\subsection{Ablation Tests}
\begin{figure}
    \centering
    \includegraphics[width=\linewidth]{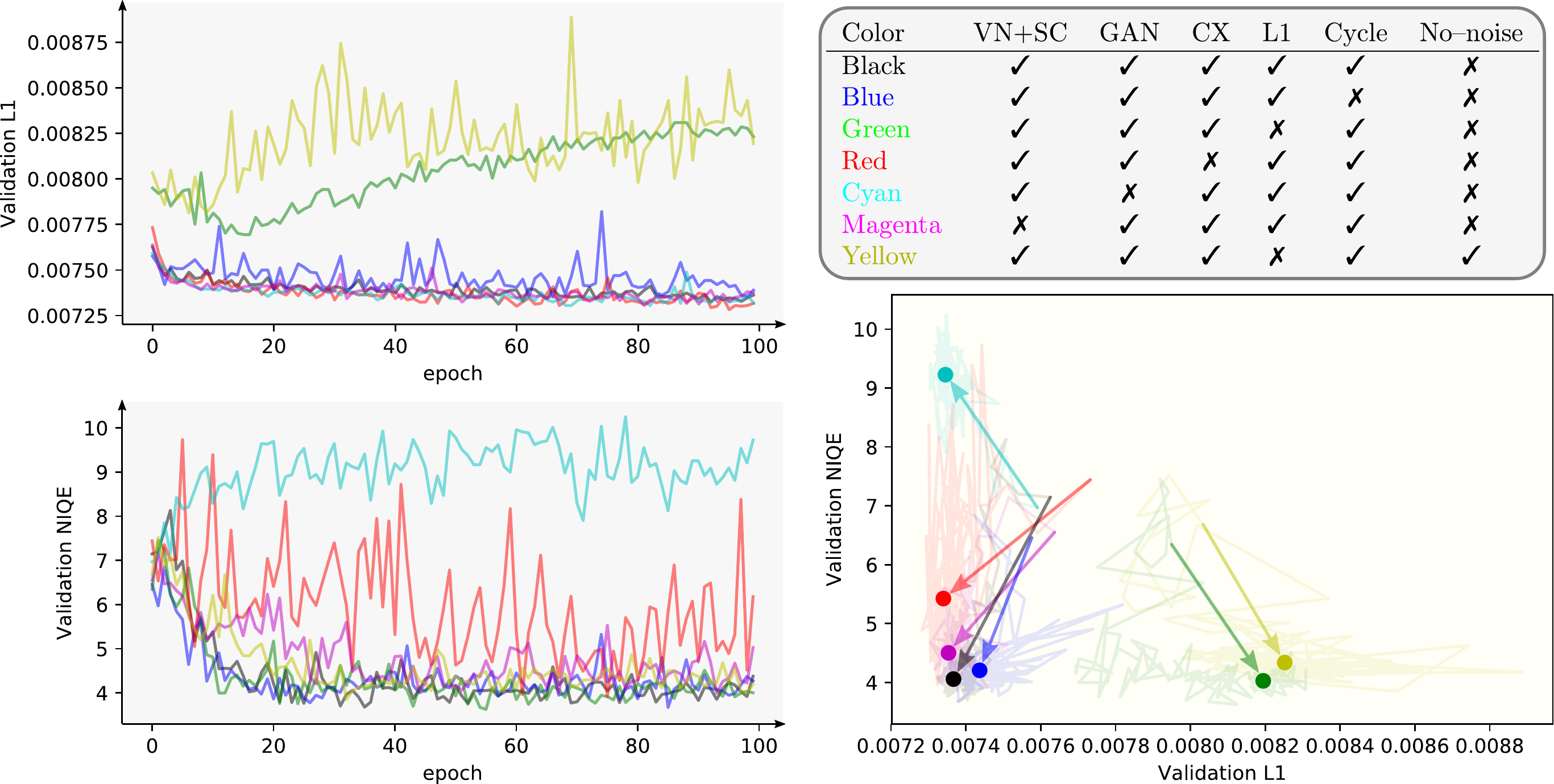}
    \caption{Ablation tests show the validation scores when training our network for $100$ epochs. We consider removal of the loss terms: GAN, CX, L1 and Cycle in \eqref{eq:total_loss}, as well as VN+SC layers in the discriminator, and training the system without noise inputs.}
    \label{fig:ablation}
\end{figure}
Our overall loss combines terms focused on different targets (e.g. low distortion, perceptual quality). In Section \ref{ssec:HF_loss} we explained the purpose of each term using the diagram in Figures \ref{fig:strategy_sets} and \ref{fig:strategy_losses}. It remains to verify this design and to quantify the relevance of each term. We also want to quantify the contribution of our novel VN+SC layer. For this purpose we trained MGBPv1 network architecture for $100$ epochs using the loss function in section \ref{ssec:HF_loss}. In Figure \ref{fig:ablation} we show our measurements of L1 (distortion) and NIQE (perceptual) in a small validation set of $14$ images after each epoch. We display the evolution through the number of epochs as well as the trajectories on the perception--distortion plane.

Overall, we see that our strategy adding all the losses (in black color) gives the best perception--distortion balance. In the extremes we see that removing the L1 and GAN losses have catastrophic effects on distortion and perception, respectively. Still, these cases do not diverge to infinity because of other loss terms. Next, it is clear that the contextual loss helps improving the perceptual quality, and regarding distortion the amount of improvement is not conclusive. Then, the addition of the cycle loss shows a clear improvement over distortion, with inconclusive improvements on perceptual quality. And finally, we observe that the addition of the VN+SC layer in the discriminator clearly improves perceptual quality, although not as much as CX and GAN losses.

Figure \ref{fig:ablation} also shows a test in which we avoid the use of noise inputs by setting $W=0$ in all losses. In this case we remove the L1 loss that would otherwise interfere with the GAN loss, causing a catastrophic effect. In this case distortion is controlled by the cycle loss, equivalent to how it is done in \cite{mechrez2018Learning}. In this configuration the network performs slightly worse in perceptual quality and clearly worse on distortion, similar to only removing the L1 loss. In this case, we believe that the network uses the randomness in the input as innovation process, which cannot be controlled and limits the diversity.

\subsection{Interpretability}
\begin{figure}
  \centering
  \includegraphics[width=\linewidth]{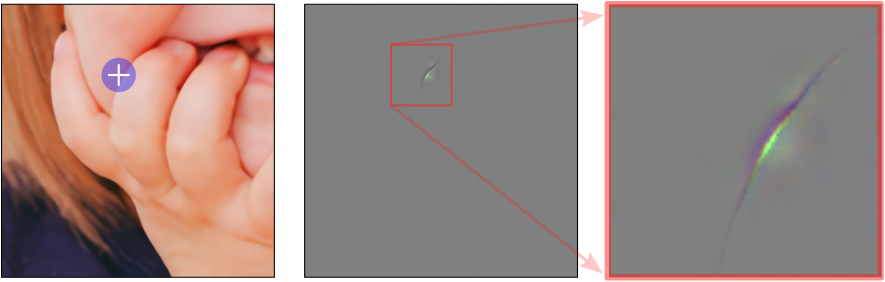}
  \includegraphics[width=\linewidth]{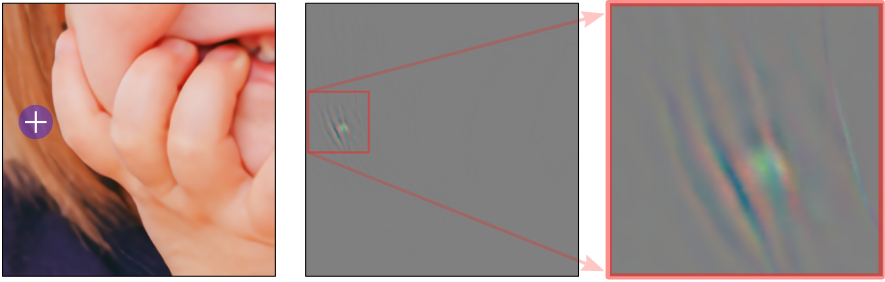}
  \includegraphics[width=\linewidth]{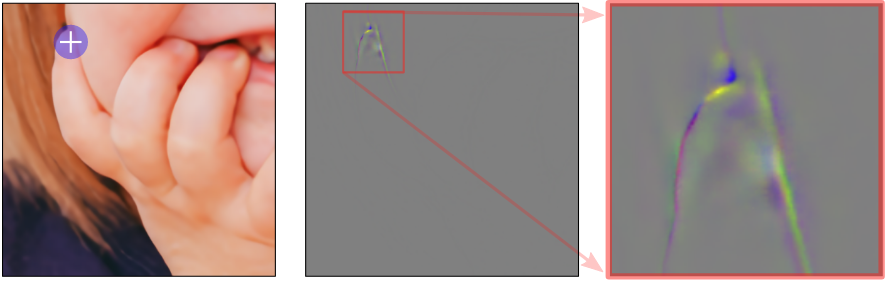}
  \includegraphics[width=\linewidth]{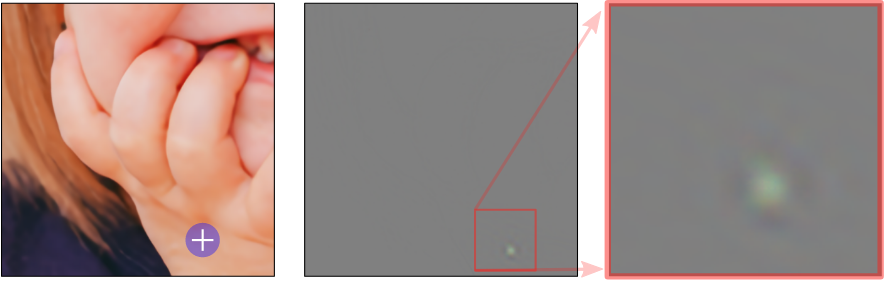}
  \caption{Deep filter visualization (DFV)\cite{PNavarrete_2019a, LinearScopes} experiments on our MGBPv2 $16\times$ model for the Fidelity track using a patch of size $767\times 767$. The model shows good knowledge of the geometry and large receptive fields. \label{fig:visualization}}
\end{figure}

Figure \ref{fig:visualization} shows interpretability results obtained by using the Deep Filter Visualization (DFV) method from \cite{PNavarrete_2019a}. To perform this complex analysis on a large model such as MGBPv2, with more than $20$ million parameters, we use the so--called \emph{Linearscope} method recently introduced in \cite{LinearScopes}. For a given pixel in the input image (blue circles on the left side), Figure \ref{fig:visualization} displays the impulse response for the network model with frozen activations (all ReLU's acting as if the input image did not change). This represents the equivalent to an upscaling filter that adapts to the pixel location. In flat areas (example at the bottom of Figure \ref{fig:visualization}) the upscaling filter looks isotropic and similar to a bicubic upscaler. In other locations, the filter strongly follows edges in hair and fingers, with receptive fields that extend for several hundred pixels. Overall, this confirms that the system has learned the geometry of the content.

\section{Conclusion}
We have proposed the MultiGrid Back--Projection (MGBP) network architecture for image super--resolution. The network combines: first, a novel type of cross--scale residual block inspired in the IBP algorithm; and second, a multigrid recursion that uses cross--scale residual blocks within cross--scale residual blocks. The result is an architecture that balances computational complexity in an efficient and effective way, focusing more on lower resolutions and decreasing computations in higher resolutions. We also introduce a particular training strategy to achieve good perceptual quality by using noise inputs to traverse the perception--distortion plane. An interpretation using the concept of innovation jumps, together with ablation tests and results on international competitions prove the effectiveness of this strategy.


%







\bibliographystyle{IEEEtran}
\bibliography{IEEEabrv,bibliography}
%



%




\begin{IEEEbiographynophoto}{Pablo Navarrete~Michelini} was born in Santiago, Chile. He received the B.Sc. in Physics (2001), B.Sc. in Electrical Engineering (2001) and the Electrical Engineer Degree (2002), from Universidad de Chile at Santiago. He received the Ph.D. degree in Electrical Engineering from Purdue University at West Lafayette, in 2008. He worked as a research intern in CIMNE at Technical University of Catalonia in 2006, and as a visitor student research collaborator at Princeton University at Princeton, NJ, in 2006--2007. He was Assistant Professor in the Department of Electrical Engineering at Universidad de Chile in 2008--2011. He worked on video processing and compression at Yuvad Technologies in 2011--2013. He joined BOE Technology Group Co., Ltd in 2014 and he is currently a Research Scientist working on machine learning methods for image and video processing applications.
\end{IEEEbiographynophoto}
\begin{IEEEbiographynophoto}{Wenbin Chen} was born in Kaifeng, China. He received a Bachelor degree in Computer Science and Technology from Xidian University, in 2016, and a Master degree in Information Technology from Melbourne University, in 2019. He majored in Embedded System at bachelor level and then transferred to Artificial Intelligence for his master degree. He joined BOE Technology Group Co., Ltd in 2019 and he works on image and video processing applications.
\end{IEEEbiographynophoto}
\begin{IEEEbiographynophoto}{Hanwen Liu} was born in Shandong province, China. He received the Bachelor degree in Automation (2010) from Central South University at Changsha, Hunan province. He received the Master degree and Ph.D. (2016) in Armament Science and Technology from Beijing Institute of Technology at Beijing. He joined BOE Technology Group Co., Ltd after graduated in 2016 and he is currently a senior researcher working on deep learning methods for image processing applications.
\end{IEEEbiographynophoto}
\begin{IEEEbiographynophoto}{Dan Zhu} received the B.E. in Communications Engineering (2014) and M.E. in Communications Engineering (2017) from JiLin University in China. She joined BOE Technology Group Co., Ltd in 2017 and is currently working on deep learning methods for image and video processing applications.
\end{IEEEbiographynophoto}
\begin{IEEEbiographynophoto}{Xingqun Jiang} received his Ph.D. degree in Electrical and Computer Engineering at Cornell University in 2009, specialized in chip technology and computer graphics. In the past 10 years, Dr. Jiang has been working in the field of IOT and artificial intelligence constantly. He joined BOE Technology Group Co., Ltd in 2015, and now serves as Chief Technology Officer in the smart system business group, primarily responsible for the product and technology innovation. In recent years, Dr. Jiang has led his team to achieve significant progress in the research and development of artificial intelligence. In 2019, they won three global championships in image super-resolution, object detection and gesture recognition, respectively. More than 20 products and solutions powered by artificial intelligence technology has been made and delivered to customers in different markets. Dr. Jiang's team has totally contributed over 2200 patents to the corporate, making BOE ranked as 6th in terms of AI patents among all the Chinese corporates.

Currently, Dr. Jiang and his team is focused on intelligent graphics and computer vision, continuing the product and technology innovation in IOT and artificial intelligence.
\end{IEEEbiographynophoto}




\end{document}